\documentclass[english]{iopart}

\usepackage{epsfig}

\begin{document}

\title[Chaotic non-attractors]
{\Large Post-critical set and non existence of preserved
meromorphic two-forms}
 
\author{ 
M. Bouamra$^\S$, S. Boukraa$^\dag$, S. Hassani$^\S$ and
J.-M. Maillard$^\ddag$}
\address{\S  Centre de Recherche Nucl\'eaire d'Alger, \\
2 Bd. Frantz Fanon, BP 399, 16000 Alger, Algeria}
\address{\dag Universit\'e de Blida, Institut d'A{\'e}ronautique,
 Blida, Algeria}
\address{\ddag\ LPTMC, Universit\'e de Paris 6, Tour 24,
 4\`eme \'etage, case 121, \\
 4 Place Jussieu, 75252 Paris Cedex 05, France} 
\ead{maillard@lptmc.jussieu.fr, maillard@lptl.jussieu.fr, sboukraa@wissal.dz, bouamrafr@yahoo.com}

\begin{abstract}

We present a family of birational transformations 
in $\, CP_2$ depending on two, or three,
parameters which does not, generically, preserve 
meromorphic two-forms. With the introduction of the orbit of the critical set (vanishing condition
of the Jacobian), also called ``post-critical set'', we get some
new structures, some "non-analytic" two-form which reduce to meromorphic
two-forms for particular subvarieties in the parameter space.
On these subvarieties, the iterates of the critical set have a polynomial
growth in the \emph{degrees of the parameters}, while one has an exponential growth out
of these subspaces.
The analysis of our birational transformation in $\, CP_2$ is first carried out
using  Diller-Favre criterion in order to find the complexity
reduction of the mapping. The integrable cases are found.
The identification between the complexity
growth and the topological entropy is, one more time, verified.
We perform plots of the post-critical set, as well as calculations
of Lyapunov exponents for many orbits, confirming
that generically no meromorphic two-form can be 
preserved for this mapping. These birational transformations 
in $\, CP_2$, which,  generically, do not preserve any meromorphic two-form, 
are extremely similar to other birational transformations we previously studied,
which do preserve meromorphic two-forms. We note that these two sets of 
 birational transformations exhibit totally similar results as far as 
 topological complexity is concerned, 
 but drastically different results as far as a
 more ``probabilistic'' approach of dynamical systems is concerned
 (Lyapunov exponents). With these examples we see that the existence of a preserved
meromorphic two-form explains most of the (numerical) 
discrepancy between the topological and probabilistic approach of dynamical systems.

\end{abstract} 
\vskip .5cm


\noindent {\bf AMS Classification scheme numbers}: 34M55, 47E05, 81Qxx, 32G34, 34Lxx, 34Mxx, 14Kxx


\vskip .1cm

\section{Introduction: Topological versus probabilistic methods in discrete
 dynamical systems}
\label{intro}

Two different approaches exist for studying discrete dynamical systems and
evaluating the complexity of a dynamical system: 
a topological approach and a probabilistic approach. A topological approach
will, for instance, calculate the topological
 entropy, the growth rate of the Arnold
complexity, or the growth rate of the successive degrees when iterating a
rational, or birational, transformation. This, quite algebraic, topological
approach is universal: one {\em counts integers}
(like some set of points, number of fixed points for the topological entropy,
number of intersection points
 for the Arnold complexity, or like the degrees of successive polynomials occurring
 in the iteration of rational
or birational transformations). This {\em universality} 
is a straight consequence
of the fact that integer counting remains invariant under any (reasonable)
reparametrization
of the dynamical system. Not surprisingly this (algebraic) topological
approach can be rephrased, or
 mathematically revisited (at least~\cite{bimero}
 in $\, CP_2$, and even~\cite{Kim} 
in $\, CP_n$), in the framework~\cite{bimero}
of a $\, H^{1,1}$ cohomology of curves in complex projective spaces
($\, CP_2$, $\, CP_1 \times CP_1$).
In this topological approach, the dynamical systems are seen as dynamical
systems of {\em complex} variables
and, in fact, {\em  complex projective spaces}. 

The probabilistic (ergodic) approach, probably
 dominant in the study of dynamical
systems, is less universal, and
amounts to describing generic orbits, introducing some (often quite abstract)
positive invariant
measures, and other related concepts like the metric entropy (integral over
a measure of
Lyapunov exponents in a Pesin's formula~\cite{Pesin}). Roughly
 speaking, we might say that a phenomenological
approach consisting in the plot of as many real 
orbits as possible (phase portraits), or in the 
calculation of as many Lyapunov exponents as
 possible, in order to get some hint of the 
``generic'' situation, also belongs to that probabilistic approach. In this
 probabilistic approach the dynamical systems 
are traditionally seen as dynamical systems
 of {\em real} variables,
dominated by {\em real functional analysis} (symbolic
 dynamics, Gevrey analyticity, ...), 
and differential geometry~\cite{Green} (diffeomorphisms, ...).

\vskip 0.2cm
\vskip 0.1cm
 
The fact that these two approaches, the ``hard'' one and the ``soft'' one, 
may provide (disturbingly)
different descriptions of dynamical systems is known by 
some mathematicians, but is hardly
mentioned in most of the graduate textbooks on discrete dynamical systems, 
which, for heuristic reasons,  try to avoid this question, implicitly 
promoting, in its most extreme form, the idea that most of the 
dynamical systems would be, up 
to strange attractors,  hyperbolic (or weakly 
hyperbolic) systems, the ``paradigm'' of 
 dynamical systems being the linearisable deterministic chaos 
of Anosov systems~\cite{Anosov,Anosov2}.  Of course, 
for such linearisable systems, these two
 approaches are equivalent. Along this line one should recall J-C. Yoccoz 
 explaining\footnote[1]{In his own address at 
the International Congress of Mathematicians in
 Zurich in 1994, or (in French) in~\cite{Yoccoz}.}
that the dynamical features that we are able to understand fall into two classes, 
hyperbolic dynamics and quasiperiodic dynamics: ``it may well happen, 
especially in the conservative case, that a system exhibits both 
hyperbolic and quasiperiodic features ...  we seek to extend these 
concepts, keeping a reasonable understanding of the dynamics, in order 
to account for as many systems as we can. The big question is then: 
Are these concepts sufficient to understand most systems'' ?  

The description of {\em conservative} cases (typically {\em area-preserving} maps, and, more generally,
mappings {\em preserving two-forms}, or $\, p$-forms) is clearly 
the difficult one, and the one for which the
distance between the two approaches, the ``hard'' one and the ``soft'' one, is 
maximum (in contrast with hyperbolic systems and, of course, linearisable Anosov systems). It is
 not outrageous to say that dynamical systems which are not
 hyperbolic (or weakly hyperbolic), or integrable (or quasiperiodic), but conservative, 
 preserving meromorphic two-forms (or $\, p$-forms), are {\em poorly understood, few tools, theorems, 
and results being available}.

In general for realistic reversible\footnote[9]{By reversible
we mean, flatly, invertible: the inverse map is well-defined, the number
of pre-image of a generic point being unique. Note that the word ``reversible'' is also used by
some authors~\cite{rob-qui-92,QuRo88} to say that the inverse 
map $\, K^{-1}$ is conjugate to the map itself $\, K$.
} mappings (which are far from being hyperbolic, 
or weakly hyperbolic, but closer to conservative systems), the equivalence of these two
 descriptions of {\em drastically different mathematical nature}
 is far from being clear.

This possible discrepancy between these two approaches (topological versus ergodic), is
well illustrated by the analysis of many discrete dynamical 
systems we have performed~\cite{BoMaRo93c,ab-an-bo-ma-2000,rearea,topo,complex}, corresponding to 
 iterations of (an extremely 
large class of) {\em birational} transformations.  These mappings
 have non-zero (degree-growth~\cite{complex} or Arnold growth rate~\cite{rearea})
 complexity,
or topological entropy~\cite{topo}, however, their orbits always look like (transcendental)
{\em curves}\footnote[2]{This
is the reason why we called these mappings ``Almost integrable'' in~\cite{BoMaRo93c}.} 
totally similar to the curves one would get with an integrable mapping,
and systematic calculations of the Lyapunov exponents of these orbits
give zero, or negative (for attracting fixed points), values.
To a great extent, the regularity of these
orbits, and, more generally,  the regularity of the whole phase portrait, seems to be related to 
the existence of preserved meromorphic two-forms (resp. $\, p$-forms)
for these birational transformations~\cite{ab-an-bo-ma-2000,rearea}. Could it
 be possible that (when being iterated) 
a birational transformation
could have a {\em non-zero topological entropy} and, in the same time, zero
 (or very small) metric (probabilistic)
 entropy, the previous ``almost-integrability'' being a consequence 
of preserved meromorphic two-forms (resp. $\, p$-forms) ?

 The existence of a preserved meromorphic two-form corresponds
 to a quite strong (almost algebraic)
 structure. Naively, one can imagine that a discrete dynamical system with a
 preserved meromorphic two-form should be ``less involved''
 than a discrete dynamical system
 without such differential structure. Should the
 existence of such exact differential structure
 be related to the ``hard'' topological, and algebraic, 
approach of discrete dynamical 
 systems (hidden K\"ahlerian structures\footnote[3]{
One may recall some exact algebraic 
(in their essence) results which are obtained
in some K\"alherian framework~\cite{Cantat,CantatFavre} (for instance,
 one inherits, immediately,
 a particular cohomology and strong differential structures~\cite{Green}). }
 for birational transformations, ...),
 or should it be related to the ``soft'' probabilistic (ergodic)  approach 
(possible relation between ``complex'' and ``real'' 
 invariant measures ...)? The answer to the previous question 
 will be fundamental to ``fill the
 gap'' between the two approaches or, at least, better
 understand the discrepancies between these two
 descriptions of birational dynamical systems. To answer this question,
 one would like to find two sets of birational
 transformations {\em as similar as possible}, but such that 
 one set {\em preserves} a meromorphic two-form, and the other
 set {\em does not preserve} a meromorphic two-form, in order
 to compare the topological and probabilistic approaches
 on these two sets. 
 
 Along this line, one should note that we found quite systematically,
 and surprisingly, preserved meromorphic
 two-forms (resp. $\, p$-forms) for an extremely large set of birational
 transformations in $\, CP_2$, and 
 in $\, CP_n$, $\, n >2$. Similar results were also found by 
 other groups\footnote[4]{J. Diller, (private communication).}
 for extremely large sets of birational transformations in $\, CP_2$.  
 Could it be possible that all birational transformations in $\, CP_2$
 preserve\footnote[5]{At first sight, such
 a strange result would present some similarity with the, still 
 quite mysterious, ``Jacobian conjecture'' of the 
 Smale's problems~\cite{Jacobconj}.} a meromorphic two-form ?
 We first need to find a first (and as simple as possible) 
 example of birational transformation in $\, CP_2$
 for which one can show, or at least get convinced of, a ``no-go'' result like
 the {\em non existence} of
 a meromorphic two-form (even very involved ...).

 The paper is organized as follows: we will first recall various ``complexity'' results 
 on a first set of birational transformations in $\, CP_2$, preserving meromorphic two-forms, 
 and we will also recall some results~\cite{bimero} of Diller and Favre
 on the topological approach of the complexity
 of these mappings. We will, then, introduce a {\em slightly}
 modified set of birational transformations in $\, CP_2$
 for which we will perform similar topological approach calculations. These calculations
 will provide, for this second set, subcases where meromorphic two-forms
 are actually preserved. This topological
 approach will yield us to introduce a fundamental tool, the {\em orbit of the critical
 set}\footnote[1]{Also called,
 by some mathematicians, {\em post critical set}, or, in short, ``PC''.
 Note that the general framework we consider here corresponds to birational
 transformations having a {\em non-empty indeterminacy set}, which is the
 natural framework when one considers {\em birational transformations} :
 the mathematician reader should forget all the theorems he knows on
 holomorphic transformations (toric monomial transformations, etc.)}, which 
 will give some strong numerical, and graphical, evidence
 that a meromorphic two-form {\em does not exist generically} 
 for this second set, outside the previous subcases. This non-existence of a  
 meromorphic two-form will be confirmed by a large set
 of Lyapunov exponents calculations, clearly exhibiting 
 non-zero positive Lyapunov exponents for this second set. We will, thus, 
 be able to conclude on the impact of the
 existence of a meromorphic two-form on the (apparent numerical) discrepancy between the
 topological and probabilistic (ergodic) approaches
 of discrete dynamical systems.

\section{Two-forms versus invariant measures}
\label{2-formversus}

Let us first recall the birational transformation $\, k_{\epsilon}\, $
 in $\, CP_2$ we have 
extensively studied from a topological (almost algebraic) viewpoint, and, also,
from a measure theory (almost probabilistic) viewpoint~\cite{ab-an-bo-ma-2000}. It is a one parameter
transformation ($ \epsilon\,  \in\, C$, or $ \epsilon\,  \in\, R$) and it reads~\cite{topo,zeta}: 
\begin{eqnarray}
\label{keps}
(x, \, y) \, \rightarrow k_{\epsilon}(x, \, y)
\, =  \, \,  (x', \, y')\, =  \, \, 
\Bigl(y \cdot {{ x+\epsilon} \over {x-1}}, \,\, x+\epsilon-1\Bigr)
\end{eqnarray}
It was found \cite{zeta} that $\, k_{\epsilon}$, the 
$\, CP_2$ birational transformation 
(\ref{keps}), preserves\footnote[2]{Birational mapping (\ref{keps})
is a particular
case of a {\em two-parameter} dependent~\cite{zeta} 
birational mapping $\, k_{\epsilon,\alpha}$, which can also
be seen to preserve a meromorphic two-form~\cite{firthcoming}.} a 
meromorphic two-form~\cite{rearea}:
\begin{eqnarray}
\label{2form}
d\mu \, = \, \,\,{{ dx \cdot dy} \over { \rho(x, \, y)}}
\,  \, = \,\, \, {{ dx \cdot dy} \over { y \, -x\, +1}} 
\end{eqnarray}
The two-form (\ref{2form}) should not be called 
a ``measure'' since the denominator
$\, y \, -x\, +1$ can be negative.
The preservation of this  two-form corresponds to the following
identity between  the covariant $\, \rho(x, \, y)\, = \, \,y \, -x\, +1$
and the Jacobian of transformation $\, k_{\epsilon}$  :
\begin{eqnarray}
\label{jac}
J(x, \, y) 
\,\, = \,\,\, {{\rho(x', \, y')} \over {\rho(x, \, y)}}
\,\, = \,\,\, {{\rho(k_{\epsilon} (x, \, y))} \over {\rho(x, \, y)}}
\end{eqnarray}
The preservation of this two-form means that this birational 
mapping can be transformed, using a (non rational) change of variables, 
into an {\em area-preserving} mapping (see page 1475
 of~\cite{rearea}, or page 391 in~\cite{ab-an-bo-ma-2000}). As
 far as a ``down-to-earth'' visualization 
of the (real) orbits, and, more generally,
of the phase portraits, is concerned,
one sees that this  $\, k_{\epsilon}$-invariant
 two-form (\ref{2form}) can actually 
be ``seen'' on the phase portrait : 
near the straight line $\,  y \, -x\, +1\, = \,0$, corresponding
 to the vanishing of the denominator of  (\ref{2form}), the points of the  phase portrait
look like a ``spray'' of points ``sprayed'' near a wall
 corresponding to this straight line (see for instance Figure 2 right,
and Figures 3, 4, 6 and 7 in~\cite{rearea}).

This birational mapping was shown~\cite{topo,zeta} to have a non-zero
topological entropy and
a degree growth complexity (or growth rate of the Arnold complexity)
associated with a quadratic number (golden number), corresponding to the 
polynomial $\, 1-t-t^2$. 
However, the extensive Lyapunov exponents
calculations we performed, systematically, gave {\em zero values} 
for all the (numerous) orbits we considered
 (see Figure 3 right, or Figures 5, 8, 10, 21, and
 pages 403 to 419 of~\cite{ab-an-bo-ma-2000}). The orbits of this mapping
 look very much like {\em curves} and,
 thus, it is not surprising 
to get zero Lyapunov exponents (see paragraphs 4 and 5
 in~\cite{ab-an-bo-ma-2000}). This
 Lyapunov exponent viewpoint, as well as the down-to-earth visualization
 of the orbits,
suggests that the mapping is ``almost an integrable mapping'', in
 contradiction with the topological viewpoint. 
Recalling, just for heuristic reasons, some 
Pesin's like formula\footnote[3]{Such a birational mapping
is not a hyperbolic system, and the various other
 birational examples we have studied
are not even quasi-hyperbolic. Pesin's formula~\cite{Pesin} (see also 
pages 299 and 400 in~\cite{ab-an-bo-ma-2000}) is
certainly not valid here. We just recall it for heuristic reasons,
just as an analogy.}, considering
the entropy as the 
integral over ``some'' invariant measure $\, d\mu_{Lyap}$ 
of the Lyapunov exponents, 
it would be natural to ask where the non zero {\em positive} Lyapunov
exponents
are hidden? Where is this apparently ``evanescent'' invariant measure of
non zero {\em positive} Lyapunov exponents? It certainly does not
correspond
to any measure describing the previously mentioned ``spray'' of points
(which could be related to
the meromorphic two-form (\ref{2form})). For invertible mappings like
birational mappings,
the known way~\cite{Smillie} of building invariant measures as
successive
 pre-images\footnote[4]{Note that, for such non invertible cases,
 we found no contradiction
between the topological approach and the probabilistic (invariant measure)
approach:
for a non-invertible deformation of (\ref{keps}) we clearly found non
zero positive Lyapunov exponents for most of the orbits
(see paragraph 8 and Figures 27 and 28 in~\cite{ab-an-bo-ma-2000}).}
 of (almost) any point, simply does not work.  Bedford and 
 Diller~\cite{BedDill} showed how to build such invariant measure
$\, d\mu_{Lyap}$ corresponding to non-zero {\em positive} Lyapunov exponents,
for the (invertible) birational transformation (\ref{keps}).
Their method amounts to considering
 two arbitrary curves\footnote[5]{They might 
even be identical.} $\, \Gamma_1$ and $\, \Gamma_2$ 
(instead of an arbitrary point), iterate $\, \Gamma_1$ with $\, k_{\epsilon}$
and $\, \Gamma_2$ with $\, k_{\epsilon}^{-1}$, and consider the limit
set obtained as the intersection of these
two different iterated curves: 
the invariant measure emerges as a wedge product
 $\, \mu^{+} \wedge \mu^{-}$.
Such a wedge product construction is actually performed
 in detail in~\cite{BedDill} on mapping (\ref{keps}).
 The invariant measure built that way,
can be seen to correspond 
to an {\em extremely slim Cantor set, which is drastically different from
the meromorphic two-form} (\ref{2form}), or, more generally, from any invariant measure
one could imagine being associated with the previously mentioned 
spray of points. 

It is also worth recalling that Bedford and Diller were
also able~\cite{BedDill} on this very example, 
but only for $\, \epsilon \, < 0$ (where only saddle points occur), 
to build some {\em symbolic dynamics coding}, yielding a $\, 2 \times 2$
 matrix that actually identifies with 
some induced pullback $\, f^{*}$ on the cohomology group\footnote[1]{See the 
cohomological approach of Diller and Favre in~\cite{bimero}, to 
get the growth rate complexity.}  $\, H^2(P^1 \times P^1)$, 
thus filling, for $\, \epsilon \, < 0$, the gap between 
a {\em real analysis} approach of dynamical systems
 and an {\em algebraic projective complex analysis} of dynamical
 systems\footnote[2]{More recently
they have been able to generalize, very nicely~\cite{DillBed},
all these results to the birational mappings $\, k_{\epsilon, \alpha}$,
depending on two parameters~\cite{topo,zeta}. Mapping (\ref{keps}) is
obtained from $\, k_{\epsilon, \alpha}$ by setting $\alpha=0$.
This mapping \cite{topo,zeta}, $\, k_{\epsilon, \alpha}$,
 can also be seen to preserve a meromorphic two-form.
Paper \cite{DillBed} provides explicit examples of a 
$\, 5 \times 5$ matrix (linear map of the Picard group), and 
a $\, 4 \times 4$ matrix, encoding the symbolic
 dynamics, such that their characteristic polynomial
both contain a factor associated with the polynomial $\, 1-t-2\, t^2-t^3$, corresponding
to the (topological) complexities of our birational family analyzed in \cite{zeta}.}.

This provides a first answer to the discrepancy between the topological 
and probabilistic approach for such birational 
transformations (\ref{keps}) (at least\footnote[3]{For
$\, \epsilon \, > \, 0$, the situation is far from being so clear.} for
$\, \epsilon \, < \, 0$): as far as 
{\em computer experiments are concerned}, the regions where the 
chaos~\cite{Shilnikov,Shilnikov2,Shilnikov3,Shilnikov4} (Smale's horseshoe, 
homoclinic tangles, ...) is hidden, is concentrated in {\em extremely narrow regions}.

\section{A first family of Noetherian mappings}
\label{Noether}

We have introduced in~\cite{Noether} a simple family of birational transformations in 
$\, CP_n$ ($n \, = \, 2, 3,  \cdots $)
 generated by the simple product
of the Hadamard inverse and (involutive) collineations. 
These birational transformations, we called Noetherian~\cite{Noether}
 mappings\footnote[4]{In reference to Noether's theorem of decomposition 
of birational transformations into products of quadratic transformations,
 like the Hadamard inverse,
and collineations~\cite{Noether}.},
present remarkable results for the growth-complexity, 
and the topological entropy,  in particular 
remarkable {\em complexity reductions}
 for some specific values of the
 parameters\footnote[5]{The parameters correspond to the entries of the collineation matrix.}
 of the mapping.
 These complexity reductions correspond
to a criterion, introduced by Diller and Favre~\cite{bimero}, based on the comparison between 
the  {\em orbit of the critical set, or even the exceptional locus},
 and the indeterminacy locus (see below (\ref{Diller})).
These mappings have similar properties compared to the
ones given for (\ref{keps}), namely a topological entropy, or 
a degree growth rate, associated with {\em algebraic numbers}, 
similar phase portraits, and the {\em existence of  preserved meromorphic two-forms}
for the transformations in 
$\, CP_2$, or, in $\, CP_n$,  preserved meromorphic $\, n$-forms, 
 together with $\, n-3$ algebraic invariants. 
In the following we will restrict ourselves to birational transformations 
in $\, CP_2$ : some of the results, we will display in the next
 sections, generalize, mutatis mutandis,
to  birational transformations in 
$\, CP_n$ ($n \, = \,  \, 3, \, 4, \, \cdots \, $) and some do not. 

\subsection{The mapping}
\label{themap}

Let us recall~\cite{Noether} the construction of the birational
 mapping $\, K$ product of a collineation $\, C$ and of a non-linear involution, 
 the Hadamard inverse,
$\, H $, acting on $\, CP_2$. We consider the standard quadratic homogeneous
transformation,  $\, H$,  
defined as follows on the three homogeneous variables
$\, (t,x,y)\, $ associated with $\, CP_2\,$: 
\begin{eqnarray}
\label{quadra}
H:\,\,\quad (t,x,y)\,\, \,\longrightarrow 
\,\,\,\,\,(x \; y,\,\, t \; y, \,\, t \; x)
\end{eqnarray}
We also introduce the following  $\, 3 \times 3\, $  matrix, acting on 
the three homogeneous variables $\, (t,x,y)\,$:
\begin{equation}
\label{C2}
M_C \, = \, \, 
\left [\begin {array}{ccc} 
a-1 &  b & c\\
a & b-1 & c\\
a & b & c-1
\end {array}\right ]
\end{equation}
and the associated collineation  $\, C$ which reads, in terms
 of the two inhomogeneous variables $\, u=\, x/t\, $ and $\, v=\, y/t$ :
\begin{eqnarray}
\label{colliC2}
&& (u, v)\, \,\, \,\longrightarrow \,\,\, \, (u', v') \,\, = \, \, \\
&&   \quad \quad \quad \quad \quad \quad \quad \quad\, = \, \,
\Bigl( {{a\, + (b-1)\, u \, + \, c\,v } \over {
(a-1)\, \, + b\, u \,+ \, c\, v}}, \, \, \, 
{{a\, + b\, u \, + \, (c-1)\,v } \over {
(a-1)\, \, + b\, u \,+ \, c\, v }} \Bigr) \nonumber 
\end{eqnarray}
The birational mapping  $\, K \, = \, C \cdot H$,
 reads, in terms of the two inhomogeneous
variables $\, u=\, x/t\, $ and $\, v=\, y/t$:
\begin{eqnarray}
\label{BiC2}
&&  K : \quad (u, v)\, \,\longrightarrow \,\, (u', v') \, = \,  \\
&&   \quad \quad \quad \quad \quad \quad \quad\, = \, \,
   \Bigl({\frac {a\, uv\, +(b-1)\, v\, +cu}
{(a-1)\, uv \, + bv\, +cu}} , \, \, \, {\frac
{a\, uv\, +bv\, +(c-1)\, u}{(a-1)\, uv\, +b \, v\, +c\, u}} 
\Bigr) \nonumber 
\end{eqnarray}

This birational mapping (\ref{BiC2}) {\em conformally}\footnote[3]{This means 
that the two-form is preserved {\em up to a constant} $\xi$.}
preserves a two-form. Actually, if one considers the product
 $\, \rho(u,v) \, = \, (u-1)\, (v-1)\,(u-v)$,
a straightforward calculation shows that 
 $\, J(u,v)$,  the Jacobian of (\ref{BiC2}),
is actually equal to:
\begin{eqnarray}
\label{equalcov} 
& & J(u,v) \, = \,\,\,\, \xi  \cdot 
 {{\rho(u',v')} \over {\rho(u,v)}}\,
\, =
\,\,\,  \xi \cdot {{ u \, v}
\over { ((a-1) \, u \, v \, + c \, u\, + \, b\, v)^3}}
\end{eqnarray}
where $\, \xi \, = \, a+b+c-1$ and where $\, (u', v')$ is the image of $\, (u, v)$
by the birational transformation (\ref{BiC2}), or equivalently
\begin{eqnarray}
\label{conf2form} 
 {{ du'\cdot dv'} \over { (u'-1)\, (v'-1)\,
(u'-v')}} \, = \,\,\,\,  \xi \cdot  {{ du\cdot dv} \over { (u-1)\, (v-1)\,
(u-v)}}
\end{eqnarray}
For  $\,\xi\, = \, 1$ (i.e. det$(M_C)=1$),
the matrix $\, M_C$, as well as its associated
collineation $\, C$, are involutions, and 
 the two-form (\ref{conf2form}) is {\em exactly preserved}. 

\subsection{Diller-Favre criterion: complexity reduction 
from the analysis of the  orbit of the exceptional locus}
\label{Diller}
We recall, in this section, the Diller-Favre method~\cite{bimero},
in order to describe the singularities of the mapping, and deduce 
{\em complexity reductions} of the mapping. In particular 
we give, for mapping (\ref{BiC2}), the equivalent 
of Lemma 9.1 and Lemma 9.2 of~\cite{bimero}.   

We assume, here, that  condition $\, c \, = \, 2-a-b\, $ is
satisfied (i.e. $\xi=1$).
 The Jacobian $\, J(u,v)\, $ vanishes on
$\, u \, = \, 0$, on  $\, v\, = \, 0$, and 
becomes infinite when 
$v\, = \, -c \, u/((a-1)\, u \, +b)$.

Using the same terminology as in~\cite{bimero},
one can show that the exceptional locus\footnote[1]{Corresponding
 to the critical set $\, J(u,v)\, = \, \,0$, together with condition
 $\, J(u,v)\, = \, \infty$.} 
of  $\, K\, $ is given by
\begin{eqnarray}
\label{exceplocus}
{\cal E}(K) & = & 
\, \,\left \{ (u=0);\, (v=0) ; \, \Bigl(v=\, {{- c \, u} 
\over { (a-1)\, u \, + b}} \Bigr) \right \}  
\end{eqnarray}
and the indeterminacy locus~\cite{bimero} of $\, K\, $ is
given by:
\begin{eqnarray}
{\cal I}(K)\, =   \,\,
 \left \{ (0,0); \,\, \Bigl( {{b}\over{(b-1)}},1 \Bigr);\, \, 
 \Bigl( 1, {{c}\over{(c-1)}} \Bigr) \right  \} \nonumber 
\end{eqnarray}
Actually, for $\, (u, \, v) \, = \, \, (0,0)$, 
the $\, u$ and $\, v$ components of $\, K$ 
are, both, of the form $\, 0/0$, for $\,(u, \, v) \, = \, \, 
( b/(b-1), \, 1)$, the  $\, v$-component of $\, K$ 
is of the form $\, 0/0$, and, for $\,(u, \, v) \, = \, \, 
(1, \, c/(c-1))$, the  $\, u$-component of $\, K$ 
is of the form $\, 0/0$.

As far as the three
 vanishing conditions (\ref{exceplocus})
 of the Jacobian, or its inverse, are concerned,
it is easy to see that their successive images
 by $\, K$ 
give respectively, when condition $\xi=1$ is satisfied:
\begin{eqnarray}
\label{0v}
&&(0, \, v) \quad \rightarrow \quad 
\Bigl( {{b-1 } \over {b}}\,, \, \, 1
\Bigr) \quad \rightarrow \quad 
 \cdots \quad \rightarrow \quad 
\Bigl( {{n\, (b-1) } \over {n\, b\, -(n-1)}}, \, 1\Bigr) \nonumber  \\
\label{0u}
&&(u, \, 0) \quad \rightarrow \quad
 \Bigl ( 1\, , \, \, {{c-1 } \over {c}}
\Bigr) \quad \rightarrow  \quad \cdots \quad \rightarrow  \quad
 \Bigl( 1, \; \; {{n (c-1)}\over{n c -(n-1)}} \Bigr) \nonumber \\
\label{ucu}
&& (u \, , \, {{- c \, u} 
\over { (a-1)\, u \, + b}}) \, \,  \rightarrow \, \, 
 \Bigl( \infty \, , \,  \infty
\Bigr)  \, \,  \rightarrow  \, \,  \cdots  \\
&& \qquad \qquad  \, \,  \rightarrow  \, \, 
 \Bigl( {{(n-1) a -(n-2)}\over{(n-1) (a-1)}} , \;\;
{{(n-1) a -(n-2)}\over{(n-1) (a-1)}} \Bigr) \nonumber
\end{eqnarray}

Do note that the iterates of ${\cal E}(K)$ for $n=\infty$ converge towards $(1,1)$
the fixed point of {\em order one} of mapping $K$.

\vskip 0.1cm

One has similar results~\cite{Noether} for  the successive images by
$\, K^{-1}$ of its exceptional locus.

At first sight it may look remarkable that 
the image  by $\, K$ of {\em curves} (like the three vanishing 
conditions (\ref{exceplocus}) of the Jacobian, or its inverse)
{\em actually blow down into points}.  
This is, in fact, a natural feature\footnote[2]{If one considers the set of 
points where the Jacobian vanishes, also called critical set, and 
assume that some
part of this critical set is not blown down into a point, then
the birational mapping would not be (locally) bijective.
Such points would have, at least, two preimages in contradiction with the
birational character of the transformation. 
This sketched proof remains valid for a birational 
transformation in $\, CP_n$ for  $\, n \, \ge 3$.}
of birational  transformations (even in $\, CP_n$). Such a phenomenon
of blow down {\em can only occur for transformations
 having a non empty indeterminacy set:}
for instance, {\em it cannot occur with holomorphic transformations}. 

One remarks that all these $\, n$-th iterates (by $\, K$ or $\,
K^{-1}$) belong (for $\, n \, \ge 2$) to the three $\, K$-invariant 
lines, namely 
$\, u\, = \, 1$,  $\, v\, = \, 1$, or $\, u \, = \, v$.  

Diller and Favre statement is that 
the  mapping $\, K$ {\em is analytically stable}~\cite{bimero} 
{\em if, and only if,} $\, \, \, K^n ( {\cal E}(K) ) \notin {\cal I}(K)\,\,  $
(respectively  $\, K^{(-n)} ( {\cal E}(K^{-1}) ) \notin {\cal I}(K^{-1})$)
for all $\, n \, \ge \,  1$. In other words the {\em complexity reduction},
which breaks the analytically stable character of the mapping,
 will correspond to situations where
some points of the orbit of the exceptional locus 
($ K^n ( {\cal E}(K) )$) encounter
the indeterminacy locus  ${\cal I}(K)$.
Having an {\em explicit} description of these orbits (see
 (\ref{0v})) for this birational transformation,
one can easily deduce the complexity reduction situations 
associated with parameters $\, a$, $\, b$, or $c$, being of the form
 $\, (N-1)/N$, where  $\, N$ 
is any positive integer.
For instance, when $\, a=\, (M-1)/M$ ($M\, $ positive integer)
 and $\, b$ generic, one gets a complexity reduction. The
complexity~\cite{Noether} being associated with 
polynomial\footnote[4]{The degree generating function~\cite{rearea,zeta}
is a rational expression with polynomial (\ref{familN}) in its denominator.} 
\begin{eqnarray}
\label{familN}
P \, = \,\,\,\,  1\,-2\,t\,+{t}^{M+1}
\end{eqnarray}
 and, similarly, 
when $\, a=\, (M-1)/M$ and $\, b\, =\, (N-1)/N$ ($M\, $
and $N\, $  positive integers),
the complexity is associated~\cite{Noether} with polynomial :
\begin{eqnarray}
\label{familMN}
P_{M,N} \, = \,\,\,\, 1\,-2\,t\,+{t}^{M+1}\,+{t}^{N+1}\,-{t}^{M+N} 
\end{eqnarray}

\subsection{Beyond the involutive condition: $\xi =a+b+c-1 \ne 1$}
\label{mune1}
Let us show that the iterates of the exceptional locus have also explicit
expressions when $\, C$ is no longer involutive (namely $\xi \ne\, 1$).
The iterates of ${\cal E}(K)$ become:
\begin{eqnarray}
\label{0vc}
&&(0, \, v) \quad  \rightarrow  \quad 
\Bigl( {{b-1 } \over {b}}, \, \, \, 1
\Bigr) \quad \rightarrow \, 
 \cdots \, \rightarrow \quad (U_n, 1) \nonumber \\
\label{0uc}
&&(u, \, 0) \quad  \rightarrow  \quad 
\Bigl ( 1,\, \, \, {{c-1 } \over {c}}\Bigr)
\quad \rightarrow \,\,
 \cdots \,\, \rightarrow \quad (1, V_n) \nonumber \\
\label{ucuc}
&&\Bigl(u,  \, \, \, {{- c \, u} \over { (a-1)\, u \, + b}}\Bigr) \, \,  
 \rightarrow  \, \, 
 \Bigl( \infty \, , \, \, \infty
\Bigr)  \, \,  \rightarrow  \, \, \, \cdots  \,\, \,  \rightarrow  \, \, 
 \Bigl( X_n, X_n  \Bigr) \nonumber
\end{eqnarray}
with:
\begin{eqnarray}
\label{Unabc}
U_n(a,b,c) & = & \, \,\, \, {\frac{(b-1)\left((a+b+c-1)^n-1\right)}
{(b-1)(a+b+c-1)^n+(a+c-1)}}  \\
V_n(a,b,c) & = & \, \, \, U_n(a,c,b), \qquad \quad  X_n(a,b,c)\, =\,\, 1/U_{n-1}(b ,a, c) \nonumber 
\end{eqnarray}

Now, the iterates of ${\cal E}(K)$ in the $n=\infty$ limit, depend on the value
of $\xi=a+b+c-1$ and read:
\begin{eqnarray}
\vert  \xi \vert  < 1 &&  \qquad U_n\,  \rightarrow \,{\frac{1-b}{a+c-1}},
\,\,\,\,\,\, V_n \,\rightarrow\, {\frac{1-c}{a+b-1}},
\, \,\,\,\,\, X_n\, \rightarrow\, {\frac{b+c-1}{1-a}} \nonumber \\
\vert  \xi \vert  > 1 &&  \qquad U_n \,\rightarrow\, 1,\,\,\,\,
\,\, \,V_n \,\rightarrow\, 1,\,\,\,\,
\, \,\, X_n \,\rightarrow\, 1
\end{eqnarray}

The above limits are precisely the fixed point(s) of {\em order
one} of mapping K which read:
\begin{eqnarray}
\Bigl(1, 1 \Bigr), \quad \Bigl(1, {\frac{1-c}{a+b-1}} \Bigr), \quad
\Bigl({\frac{1-b}{a+c-1}}, 1 \Bigr), \quad
\Bigl({\frac{b+c-1}{1-a}}, {\frac{b+c-1}{1-a}} \Bigr) \nonumber
\end{eqnarray}

Again, one remarks that all these $\, n$-th iterates (by $\, K$ or $\,
K^{-1}$) belong (for $\, n \, \ge 2$) to the three $\, K$-invariant 
lines $\, u\, = \, 1$,  $\, v\, = \, 1$, or $\, u \, = \, v$, allowing 
a meromorphic two-form like (\ref{conf2form}) to be (conformally) preserved.  

For $\xi=1$, the four fixed points of  order one collapse to a only one.
For $\xi \ne 1$, the iterates of the exceptional locus converge to one, or more
than one, fixed point(s) of order one.

\section{A second family of Noetherian mappings}
\label{antisto}
Let us, now, introduce another set of birational transformations in $\, CP_2$,
built in a totally similar way as the Noetherian mappings~\cite{Noether}
of the previous section, 
namely as product of a collineation $C$ and the previous quadratic
transformation
  $H$ (Hadamard inverse (\ref{quadra})). Our only 
slight modification is that the $\, 3 \times 3$ matrix
 $M_C$, associated with this collineation, is now the {\em transpose}
 of matrix $M_C$
considered in~\cite{Noether} and previously given in (\ref{C2}). It is
straightforward to remark
that $\, \xi \, = \, a+b+c-1\, = \, 1$ is, again, the condition for
collineation $\, C$ to be an involution ($\det(M_C) =1$).
In that involutive case it is also straightforward to see that $\,K^N$,
and  $\,K^{-N}$, are conjugated :  $\,K^{-N} \, = \, \,$
$ C \cdot K^N \cdot  C $ $\,\, = \,C^{-1} \cdot K^N \cdot  C \,$
 $\,\, = \,H^{-1} \cdot K^N \cdot  H \,$
 $\,\, = \,H \cdot K^N \cdot  H$.
Thus transformations $\,K$ and  $\,K^{-1}$ have {\em necessarily
the same complexity}.    Most of the results
we will display in the following, will be restricted (for heuristic reasons) 
to this involutive condition 
$\, \xi \, = \, a+b+c-1\, = \, 1$, but it is important to keep 
in mind that many of these results can be 
generalized to the non-involutive case $\, \xi \, \ne \, 1$. 

The mapping $\, K\,=\,C \cdot H$, in terms of inhomogeneous
variables ($u=x/t$, $v=y/t$), reads:
\begin{eqnarray}
\label{defKanti}
&& K: \,\, (u,v)
\,\,\longrightarrow \,\,  \\
&&  \quad \quad \quad \quad  \Bigl( {\frac{b\, uv\,+(b-1)\, v\,+b\, u} 
 {(a-1) \, uv\, +a\, (u+v)}},\,\,
{\frac{c \, uv \, + c\, v \, +(c-1) \, u}
{(a-1) \, uv +a\, (u+v) }} \Bigr) \nonumber
\end{eqnarray}

When written in a homogeneous way, it is clear, since the three 
homogeneous  variables, as well as 
the three  parameters  $\, (a, b, c)$, 
 are on the same footing, that transformation 
$\,K\, =\,C \cdot H$ must exhibit a symmetry with respect to the
group of permutations of the three (homogeneous) variables.
The symmetry, induced by this group of permutations of the
 three homogeneous variables,
leads to equivalence between mappings with different couple of 
parameters $a$ and $b$ (with $c=2-a-b$). 
The change $(a,b) \rightarrow (b,a)$ combined with
$(u,v) \rightarrow (1/u, v/u)$, and the change
$(a,b) \rightarrow (a, 2-a-b)$
combined with $(u,v) \rightarrow (v,u)$, leave the mapping $K$ unchanged.
Defining the two involutions
\begin{eqnarray}
P: (a,b) \,  \longrightarrow \,  (a,2-a-b), \qquad \qquad 
T: (a,b) \,  \longrightarrow \,  (b,a)  
\end{eqnarray}
the parameter plane $(a,b)$ is composed of six
 equivalent regions reached by five
transformations of one region. 
The five regions are reached from (e.g.) the region 
$\, 1\, -a/2\,  \le\,  b \, \le\,  a$
by the action of\footnote[3]{Note that $P\cdot T $ (or $ T\cdot  P$)
 is an order three symmetry.
} $P$, $T$, $P\cdot  T$, $T\cdot  P$ and $P \cdot T\cdot  P$.
It means that the mappings built with one of the matrices
$M_C$, $P\cdot M_C$, $T \cdot M_C$, $P\cdot T \cdot M_C$, $T\cdot P \cdot M_C$,
$P\cdot T\cdot P \cdot M_C$ are equivalent. As a consequence, if $(a, b)$
 gives the complexity
$\lambda$, so do $P(a, b)$, $T(a, b)$,
$P\cdot T(a, b)$, $T\cdot P(a, b)$, $P\cdot T\cdot P(a, b)$ for
the corresponding mapping.
The fixed points of the involutions $P$, $T$ and $\, P\cdot T\cdot P$ lie,
respectively, on three lines:
\begin{eqnarray}
\label{ablines}
  b=\, 1-a/2, \qquad \quad  b=\, a, \qquad \quad   b=\, 2-2a  
\end{eqnarray}
These three lines present interesting properties as will be seen in the 
following.
The fixed point of $\,P \cdot T$, or $\, T \cdot P$, 
correspond to a point $\, a=b=2/3\, $
in the $\, (a, \, b)$ parameter plane (we will see below that 
 this corresponds to an integrable mapping).  
As far as symmetries in the $\, (a, \, b)$ parameter plane are concerned,
another codimension-one subvariety pops out, namely the quadric
\begin{eqnarray}
\label{defC0}
C_0(a, \, b)\, =\,\,\, a^2\, +b^2\, +a b\, -2 (a\, +b)\, =\,\, 0
\end{eqnarray}
which is invariant under the five transformations
$P$, $T$, $P\cdot T$, $T\cdot  P$ and $P \cdot T \cdot P$.
Having a genus 0, curve (\ref{defC0}) has a rational parametrization.

Condition $C_0(a,b)=0$ occurs as a condition for $\, K$ {\em to be an 
order two} transformation not in the whole $(u, \, v)$ plane, but {\em on some singled-out 
curve} (see the algebraic curve (\ref{list2}) below).  Note that, an algebraic curve such that 
$\, K^2(u, \, v) \, = \, \, (u, \, v)$ is {\em necessarily a
 covariant} curve for $\, K$.

\subsection{Diller-Favre complexity reduction analysis on
 the new Noetherian mappings}
\label{complexanaly}
In order to perform a complexity reduction analysis on (\ref{defKanti}),
similar to the one displayed in section (\ref{Diller}),
 based on the Diller-Favre criterion,
 let us calculate the  Jacobian of $\, K$, the birational 
transformation (\ref{defKanti}):
\begin{eqnarray}
\label{jact}
J(u, \, v)\, =\,\,\,\, 
 {\frac { \left( a+b+c-1 \right)\, uv  }
{ \left((a-1)\,uv\,+a(u+v) \right) ^{3}}}
\end{eqnarray}
Denoting $J^{(-1)}$ the Jacobian of
$K^{-1}$,
one  easily verifies that (as it should):
\begin{eqnarray}
 J(K^{-1}(u, \, v))  \cdot J^{(-1)}(u, \, v)\, \,  = \, \ \, 
 J(u, \, v)  \cdot J^{(-1)}(K(u, \, v)) \,\,  =\,\, + 1 \nonumber 
\end{eqnarray}

The finite set of points of indeterminacy of the mapping, ${\cal I}(K)$, and  
the finite set of exceptional points of the mapping (critical set $\, J=0$, 
together with condition
$\, J=\infty$), ${\cal E}(K)$,
read: 
\begin{eqnarray}
{\cal I}(K) & = & \,\, \left \{ I_1, I_2, I_3 \right \}\,=\,\,
\left \{ (0,0); ({{ a} \over {b}}, {\frac{a}{1-a-b}});({\frac{a}{b-1}}, 
{\frac{a}{2-a-b}}) \right \} \nonumber \\
{\cal E}(K) & = \,\, & \left \{ V_1, V_2, V_3 \right \}\,=\,\,
\left \{ (u=0); (v=0);(u={\frac{-av}{v(a-1)+a}}) \right \} \nonumber 
\end{eqnarray}
Let us focus on the first iterates of one of the three vanishing conditions
of the  Jacobian $\, V_2$, namely $\, v \, = 0$:
\begin{eqnarray}
\label{singsing}
&& (u_1, v_1) \, =\, \, \, 
\Bigl( {\frac{b}{a}},\,\, {\frac{1-a-b}{a}}\Bigr), \nonumber \\
&& (u_2, v_2) \, =\, \, \,
\Bigl({\frac{(b-1)}{(a-1)}}{\frac{(C_2^{22}+b)}{(C_2^{22}+a)}},\,\,
{\frac{(1-a-b)}{(a-1)}}
{\frac{(C_2^{22}-a-b)}{(C_2^{22}+a)}}\Bigr), \nonumber \\
&& (u_3, v_3) \, =\, \, \, \cdots 
\end{eqnarray}
The expression $C_2^{22}$ is given in (\ref{C222}) below.
Do note that, in contrast with the situation encountered
in the previous section (see (\ref{0v}), (\ref{Unabc})), the degree growth 
of (the numerator or denominator of) these successive expressions
in the parameters $a$ and $b$ is, now, {\em actually exponential, and, thus, 
one does not expect closed forms for the successive iterates} ($u_N$, $v_N$).
We will denote $\, \delta$ the degree growth rate (complexity)
 associated with the exponential degree growth
 $\, \simeq  \delta^N$ of these $u_N$'s 
and $v_N$'s (in the $(a, \, b)$ parameters).
 This degree growth rate (in the parameters $a$ and $b$)
 of the iterates of the vanishing
 conditions of the Jacobian depends on the values of $a$ and $b$.
 In the previous section (see (\ref{0v}), (\ref{Unabc}))
this degree growth rate was $\delta=1$ for generic values of the parameters.

Before performing any calculation, let us remark that, due to the
 previously mentioned permutation symmetry,
the nine ``Diller-Favre conditions''  $K^N( {\cal E}(K) ) \in I(K)$ 
for complexity reduction, are related
\begin{eqnarray}
\label{equiv}
& & K(V_1) \in I_1 \, \Longleftrightarrow \,  P \cdot K(V_2) \in I_1, 
 \quad \,\,
K(V_1) \in I_2  \, \Longleftrightarrow \,  K(V_2) \in I_3,  
\nonumber \\
& & K(V_2) \in I_2  \, \Longleftrightarrow \,  P \cdot K(V_1) \in I_3, 
 \quad \,\,
K(V_3) \in I_3  \, \Longleftrightarrow \,  P \cdot K(V_3) \in I_2 
\nonumber
\end{eqnarray}

The method in~\cite{bimero} amounts to solving $K^N(V_i) \in I_j$.
One obtains, for mapping (\ref{defKanti}),
 algebraic curves in the $(a, \, b)$-plane, with some singled-out
 $(a, \, b)$ points. These
algebraic curves appear, at {\em even} orders,
as common polynomials (gcd) in the components
of $K^N(V_1) \in I_3$, or  $K^N(V_2) \in I_2$ or
 $K^N(V_3) \in I_1$.
Let us call these algebraic curves associated with conditions 
$K^N(V_i) \in I_j$, respectively
$C_N^{13}$, $C_N^{22}$ and $C_N^{31}$ ($N$ being even).
For instance $C_2^{22}$ corresponds to $K^N(V_2) \in I_2$,
 that is $\, (u_2, \, v_2)\, = \, (a/b, \, a/(1-a-b))$, which reads 
 $\, (a^2+ab+b^2)-(a+b)\, = \,0$. 
These algebraic curves are $(a, \, b)$-subvarieties of complexity growth,
for (\ref{defKanti}),
 lower than the generic one ($\lambda=2$), and they are related
by $\,  P \cdot C_N^{13} = C_N^{22}$ and $\,  T \cdot C_N^{13} = C_N^{31}$.
They are polynomials in $a,b$ of degrees $2$, $6$, $12$, $26$,
$48$, $98$, $\cdots\, $ (for $N \, = \, 2, \, 4, \, 6, \, 8, \, 10, \, 12, \cdots$).
Since they are calculated from the $\, u_N$'s  and 
$\, v_N$'s (\ref{singsing}) which are rational expressions
in $(a, \, b)$ with corresponding polynomials
of degree growing exponentially like $\, \delta^N \sim 2^N$, it is not 
surprising to see the degree of these successive $(a, \, b)$ polynomials
growing exponentially, but with a {\em lower} rate (see Appendix A).

Note that the singularities of these algebraic curves
(from a purely algebraic geometry viewpoint: local branches, ...)
correspond to  points $(a, \, b)$, in the parameters
space, for which the birational transformation $\, K$ has actually {\em lower}
complexities (see Appendix A). 
Note that the singularities of the curves $\, C_N$'s contain those of
the curves of lower $N$. A detailed analysis of this set
of curves, their mutual intersections, and the relation between these
intersections, and  singled-out (singular) points of the curves, and
the associated further reduction of complexity, will not be performed here.

The polynomials $\,C_N^{22}$ appearing in this complexity reduction analysis,
are, of course, symmetric in $\, a$ and $\,b$.
Those of the first orders read:
\begin{eqnarray}
\label{C222}
&& C_2^{22}  = \, (a^2+ab+b^2)-(a+b)  \\
&& C_4^{22} = \,(a^2+ab+b^2)^3
-(a+b)(a^2+ab+b^2)(4a^2+7ab+4b^2) \nonumber \\
&& \qquad  \quad  +(7a^4+26a^3b+36a^2b^2+26ab^3+7b^4) \nonumber \\
&& \qquad  \quad  -(a+b)(6a^2+11ab+6b^2)+(2a^2+3ab+2b^2) \nonumber 
\end{eqnarray}
These polynomials  $\,C_{N}^{ij}$ ($ij=13,22,31$)
have been obtained up to  $\, N\, = \, 12$.
Some of their algebraic geometry properties (singularities, genus, ...) 
are summarized in Appendix A.
\begin{figure}
\psfig{file=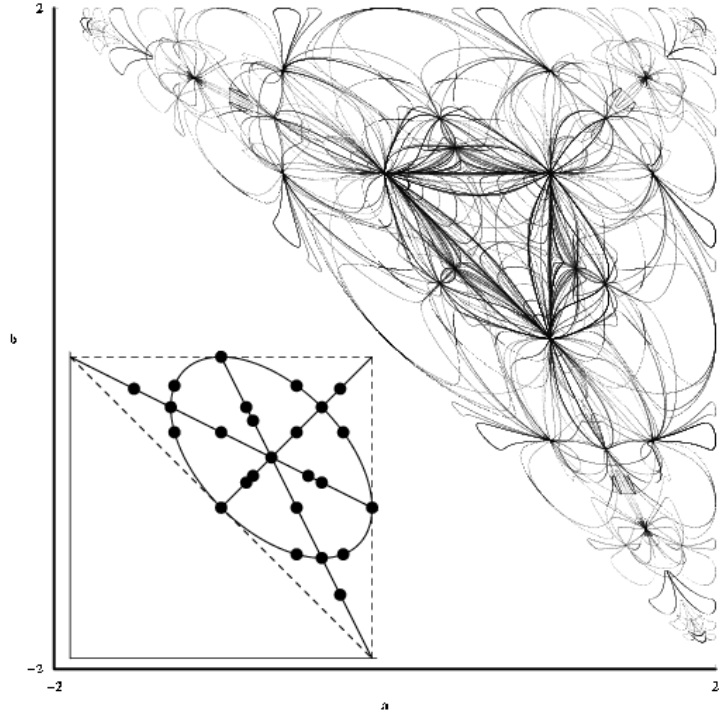,scale=0.53}
\caption{Polynomials $C_N$ in the $(a, \, b)$ parameter plane (upper right corner).}
\label{f:fig1}
\end{figure}

Let us display these various algebraic curves
 $\, C_N^{ij}$ in the $(a, \,\, b)$-parameter plane. One sees, on Figure \ref{f:fig1} (upper right corner), 
that this accumulation of curves
looks, a little bit, like a (discrete) ``foliation'' of the
$(a, \,\,b)$-plane in curves similar to a linear pencil
 of algebraic curves~\cite{Cremona}, the 
``base points'' of this linear pencil being, in fact, singular points
of these $\, C_N$'s (see Appendix A) of lower 
complexity and sometimes, $(a, \,\,b)$ points for which the 
mapping becomes integrable.

On these algebraic  curves
$\,C_N^{ij}\, = \, 0$ $(N=2, 4, 6, \cdots)$, the
 complexity is given by the inverse of
the smallest root of:
\begin{eqnarray}
\label{shift}
1-2t+t^{N+2}\, =\, \, 0
\end{eqnarray}
As $N$ increases, the complexity reads $\lambda=1.8392, 1.9659,$
$  1.9919, \cdots $
One recovers a family of complexities (depending on $\, N$) 
already seen for the Noetherian mappings~\cite{Noether}
of the previous section and, 
even,  for the  
mapping (\ref{keps}) for $\, \epsilon \, = \, 1/N$ (see~\cite{zeta}). 
Actually, one finds a shift of $\, +1$
 between (\ref{familN}) and (\ref{shift}).

In contrast with the situation encountered with the
 Noetherian mappings of the previous section (\ref{Diller}) (see also~\cite{Noether}),
the complexity reduction conditions are now involved families
 of polynomials (exponential degree growth in the $(a, \, b)$
 parameters i.e. $\delta >1$),
 instead of the previous extremely simple, and 
{\em separated} conditions~\cite{Noether} 
 in the $\, a$, $\, b$, $\,c$ 
variables ($a \, = \, (N-1)/N$, ...). 

Recalling the complexity reduction scheme described in 
section (\ref{Diller}) for mapping (\ref{BiC2}),
we saw further complexity reductions on the intersections of
two complexity reduction conditions $\, a=\, (M-1)/M$ and 
$\, b\, =\, (N-1)/N$ ($M\, $
and $N\, $  positive integers)
and $\, c \, = \, 2-a-b\, $,
namely families of complexities depending on the two integers
$N\, $ and $\,M$
 associated~\cite{Noether} with polynomials 
$\,1-2\,t+{t}^{M+1}+{t}^{N+1}-{t}^{M+N}$. 

By analogy, it is natural to see if a similar complexity reduction scheme
also occurs for mapping (\ref{defKanti}), by
calculating the degree growth complexity when the parameters $a$, and $b$, 
are restricted to the intersection of two conditions  $\,C_N^{ij}\, = \, 0$. 
Actually, we have considered the intersection of
 $\, C_2^{31}\, = \,0\, $ and $\, C_4^{22}\, = \, 0$,
that we will denote symbolically 
$\,  C_2^{31} \cap  C_4^{22}$, as well as the intersection
 $\,  C_2^{13} \cap  C_4^{31}$.
We obtained the following
generating function in agreement with the successive degrees (up to $t^9$)
in the corresponding iteration:
\begin{eqnarray}
\label{gener2}
&& G_{C_2^{13} \cap  C_4^{31}} \, = \, \,1+2\,t+4\,{t}^{2}
+7\,{t}^{3}+13\,{t}^{4}
+24\,{t}^{5}+43\,{t}^{6}+77\,{t}^{7}\nonumber \\
&& \qquad \quad \quad +138\,{t}^{8}+247\,{t}^{9}
\, + \cdots   \, = \, \,{\frac 
{1-{t}^{3}}{1-2\,t+{t}^{4}+{t}^{6}-{t}^{8}}}
\end{eqnarray}

Keeping in mind the shift of $\, +1$
between (\ref{familN}) and (\ref{shift}), one might expect a
formula like (\ref{familMN}) for
an intersection $\, C_M^{13} \cap  C_N^{31}$ (or $\, C_M^{31} \cap  
C_N^{22}$)
\begin{eqnarray}
\label{familMNbis}
Q_{M,N} \, = \,\,\, \,  1\,-2\,t\,+{t}^{M+2}\,+{t}^{N+2}\,-{t}^{M+N+2} 
\end{eqnarray}
This is actually the case with the  previous example
(\ref{gener2})
where one has $\, M\, = 2$ and $\, N \, = \, 4$. Another example, also in agreement with
(\ref{familMNbis}),  corresponds to the intersection
 $\,  C_4^{31} \cap  C_6^{13}$ 
for which one gets a rational degree generating function with 
denominator
$1-2\,t+{t}^{6}+{t}^{8}-{t}^{12}$.

Note that such formula seems to remain valid even when $\, M \, = \, N$. 
For instance,
for $C_4^{13} \cap  C_4^{31}$ the denominator 
of the generating function reads $\, 1-2\,t+2\,{t}^{6}-{t}^{10}$,
and for $C_6^{13} \cap  C_6^{31}$ the denominator 
reads $\,1-2\,t+2\,{t}^{8}-{t}^{14}$,
in agreement with (\ref{familMNbis}) for $\, M \, = \, N\, = \, 6$. 

One sees that one has exactly the same complexity reduction scheme, and 
{\em the same family of complexity},
as the one depicted in Section (\ref{Diller}) for~\cite{Noether}.

However, one does see a difference with the
 intersection of three conditions.
For mapping (\ref{BiC2}), we saw~\cite{Noether}
that the intersection of three conditions
$\, a=\, (N-1)/N$, $\, b=\, (M-1)/M$, $\, c \, = 2-(a+b)=\, P/(P+1)$, 
yields
systematically {\em integrable} mappings.
Here the $(a, \, b)$ points
 corresponding to  intersection of three conditions
 $\,C_N^{ij}\, = \, 0$ when they exist,  may still yield an 
exponential growth of the calculations of lower complexity:
\begin{eqnarray}
\label{gener7}
&& G_{C_6^{13} \cap  C_6^{31} \cap C_8^{22} } \, = \, \,
1+2\,t+4\,{t}^{2}+8\,{t}^{3}+14\,{t}^{4}+24\,{t}^{5}+40\,{t}^{6}+66\,{
t}^{7} \nonumber \\
&& \qquad \quad \, +108\,{t}^{8}
\, + \cdots \,  \, = \, \,{\frac {1+{t}^{3}}{ \left( 1-t \right)  
\left( 1-t-{t}^{2} \right) }}
\, = \, \,{\frac {1+{t}^{3}}{1-2\,t+{t}^{3}}}
\end{eqnarray}
We have a similar result for the intersection of the three curves 
$\, C_2^{31} \cap  C_6^{31} \cap C_{10}^{31}$
with a denominator reading $\,1-2\,t+{t}^{4}$.

One should remark, in contrast with most of the
degree growth rate calculations we have performed for so
many birational transformations~\cite{complex}, that one can hardly 
find rational values for the two parameters $\, a$ and $\, b$, 
lying on the various $\, C_N^{ij}$'s we have just considered,
(and of course it is even harder for intersections of such algebraic curves), 
such that one would deal with iterations of birational transformations with
integer coefficients,
and factorization of polynomials with integer coefficients. 
Such $\, (a,\,  b)$ points on $\, C_N^{ij}$ algebraic curves or 
intersections of such curves, are {\em algebraic numbers}. 
The degrees of the successive iterates should correspond
to factorizations performed in some field extension
corresponding to these algebraic numbers and curves. In practice, results and
series like the ones displayed above ((\ref{shift}), ..., (\ref{gener7})),
cannot be obtained this way.
To achieve these factorizations, we have introduced a ``floating''
factorization method that is described in Appendix B.

\subsection{Degree growth complexity versus topological entropy }
\label{topol}

The topological entropy is related to 
the growth rate of the number of fixed
points of $\, K^N$ (see~\cite{rearea}).
The counting of the number of primitive cycles of order $\, N$,
for the generic case $[4, 1, 2, 3, 6, 9, 18, 30, \cdots \,\, ]$
gives a {\em rational} dynamical zeta function~\cite{zeta}
\begin{eqnarray}
\label{generic}
 \zeta_g(t) \, = \, \, \,     {{ 1} \over {\left( 1-2\,t \right)
 \left( 1-t \right) ^{2} }} 
\end{eqnarray}
which is related to the homogeneous degree generating  
function $G(K)$ by the identity:
\begin{eqnarray}
\label{rela1}
{\frac{t}{\zeta_g}} \cdot {{d} \over {dt}} \zeta_g  \,\,  = \, \,  \, \,   
2\, G(K)(t) \, + \, \, {\frac {2\, t}{1-t}}
 \,\,  = \, \,\,   \, {\frac {2\,t}{1-2\,t}}   + \, \,{\frac {2\, t}{1-t}}
\end{eqnarray}

Restricted to the curve of complexity reduction
$\, C_2^{22}(a, \, b) \, =\, 0$, the primitive fixed points become
$[4, 1, 2, 2, 4, 5, 10, 15, 26, 42, \cdots\,\,]$ giving the rational dynamical zeta function:
\begin{eqnarray}
\label{C2zeta}
 \zeta(t) \, = \, \, {\frac {1}{ \left(1-2\,t+{t}^{4} \right)  
\left(1-t \right) ^{2}}}  
\end{eqnarray}
Again, note that this  dynamical zeta function is related to the homogeneous degree 
generating  function $G(K)$
(corresponding to $\, C_2^{22}(a, \, b) \, =\, 0$), by the identity:
\begin{eqnarray}
\label{rela2}
{\frac{t}{\zeta}} \cdot {{d} \over {dt}} \zeta \,  \, = \, \,  \, \, 
2\,G(K)(t) \, + \, \, {\frac {2\, t}{1-t}} \,\,\, 
= \, \, \, 
 {{ 2\, t \cdot (1-2\, t^3) } \over { 1-2\,t+{t}^{4}}}
 \, + \, \, {\frac {2\, t}{1-t}} 
\end{eqnarray}

We thus see, with these two examples (and similarly to the results
 obtained for the birational transformations~\cite{rearea,zeta}
as well as the Noetherian mappings~\cite{Noether}), an {\em 
identification
 between the growth rate  of the number of fixed
points of} $\, K^N$,  {\em and the growth rate of the degree of
 the iteration} (previously studied (\ref{defKanti})), or
 equivalently, the {\em  growth rate of the Arnold complexity}.  

Relations (\ref{rela1}) and (\ref{rela2}) are in agreement with a
Lefschetz formula\footnote[4]{
The Lefschetz formula is well defined 
in the {\em holomorphic} framework (see page 419 in~\cite{Griffiths}),
but  is much more problematic in the {\em non-holomorphic 
case of birational transformations
for which indeterminacy points take place}: in very simple words one could say
that, in the Lefschetz formula, 
 some fixed points are ``destroyed'' by the indeterminacy points.
 A good reference is~\cite{Fourier}.}:
\begin{eqnarray}
\label{mudd}
  \nu_N \, = \, \, \, d_N(K) \, + \, \,  d_N(K^{-1}) \, + \, \, 1 \, + \, \, 1
\end{eqnarray}
where $ \, \nu_N$ denotes the number of fixed points of $ \,K$ or  $ \,K^{-1}$,
  $ \,d_N(K)$ denotes the degree of  $ \,K^N$,  $ \,d_N(K^{-1})$ the degree  $ \,K^{-N}$.
This formula (\ref{mudd}) means that the  number of fixed points 
is the sum of four ``dynamical degrees \cite{Fourier}'' 
$ \, \delta_0 \, + \,  \delta_1 \, + \,  \delta_2 \, + \, 
 \delta_3$. Dynamical degree $ \, \delta_0 $ is always equal to  $ \,+1$,  $ \, \delta_3 $ is
the topological degree (number of preimages:  $ \, \delta_3 $ is equal to  $ \,+1$
for a birational mapping), $ \, \delta_1 $ is the first dynamical degree (corresponding to
 $ \,d_N(K)$) and $ \, \delta_2 $ is the second dynamical degree (corresponding to
 $ \,d_N(K^{-1})$). 
\vskip .1cm

{\bf Remark 1:} Most of the physicists will certainly take for granted that
the degree growth rate corresponding to the iteration of $\, K$ and its inverse $\, K^{-1}$
identify: $ \,d_N(K) \simeq \, \lambda(K)^N,$ $ \, d_N(K^{-1}) \simeq \, \lambda(K^{-1})^N$,
with $\,  \lambda(K) \, = \, \lambda(K^{-1})$. This is actually the case
for all the birational transformations we have studied~\cite{zeta}.
In the specific examples of this paper,
this is, in the involutive case $\, \xi\, = \,a+b+c-1\,  = \,1$,
 a straight consequence of the fact that $\, K$ and $\, K^{-1}$ are conjugated.
 More generally,
this fact can be proved for all birational transformations in $\, CP_2$,
but certainly not
for birational transformations in 
$\, CP_n$, $\, n \, \ge 3$ (for instance birational transformations
generated by products of more than two
 involutions, or ``Noetherian'' mappings products of many
collineations and Hadamard involutions~\cite{Noether}, such 
that  $\, K$ and $\, K^{-1}$ are not conjugated).
Appendix C provides a simple example of bi-polynomial transformation in $CP_3$ 
such that $\,  \lambda(K) \, \ne  \, \lambda(K^{-1})$.
\vskip .1cm 

{\bf Remark 2:} The very definition of the dynamical zeta function on
$\, C_0(a, \, b)\, = \, 0$ is  a bit subtle, and problematic,  since the number
of fixed points for $\, K^2$ (and thus $\, K^{2N}$) is actually {\em infinite} (one has 
a {\em whole curve} (\ref{list2}) of {\em fixed points of order two}).
Apparently, in that case where an infinite number of fixed
 points of order two exist, one does not seem, beyond these
cycles of order two,  to have primitive  cycles of even order.
Introducing the dynamical zeta function as usual, from
an infinite Weil product~\cite{zeta}
on the cycles, and
taking into account just the odd cycles, 
one obtains (more details are given in Appendix D)  
that this zeta function verifies a simple functional
equation
\begin{eqnarray}
  \zeta(t^2) \, = \, \,
 {{(1-2\,t) \cdot (1-t)^2} \over {(1+2\,t) \cdot (1+t)^2 }}
 \cdot \zeta(t)^2
\end{eqnarray}
showing that, the complexity is still the generic
$\, \lambda \, = \,2$ but,
this time, with  an expression
 which {\em  is not a rational function, 
but some ``transcendental''expression}. 
In order to have a Lefschetz formula (\ref{mudd}) remaining  
valid, in such highly singled-out cases for dynamical zeta functions,
one needs to modify the definition
of the dynamical zeta function so that it is no longer 
deduced from an infinite Weil product~\cite{zeta} formula
on the {\em cycles}. To be more specific, this must be performed using 
the so-called~\cite{Fulton} ``Intersection Theory'' which is a
(quite involved) theory introduced to cope with isolated points, {\em as well
as non-isolated points} (curves ...), introducing
some well-suited (and subtle) concepts like the notion of {\em multiplicity}.
All the associated counting of intersection numbers will, then, correspond
to counting
of {\em finite integers} (replacing the counting of cycles ...).
This is far beyond the scope of this very paper. 

\section{Preserved meromorphic two-forms in particular subspaces $(a,b)$}
\label{covar}

In Appendix E, we show that the degree growth (in the $(a, \, b)$ parameters) for
the iterates of the \emph{three curves of the critical set} (resp.
exceptional locus)
when the parameters are restricted to $b=a$, $b=2-2a$, $b=\, 1-2a$,
and $C_0(a,b)=0$, is {\em polynomial} ($\delta=1$).
The iterates are found in closed expressions.
Let us show that, in these cases, the mapping $K$ 
preserves simple meromorphic two-forms.

On the three lines $\, b=a$, $b=\, 2\, -2a\,$, and $ b=1-a/2$,
 one finds three preserved meromorphic
 two-forms reading respectively:
\begin{eqnarray}
\label{list}
&& {{ du \cdot dv} \over
 { (u-1) \cdot ( 2\, (2a-1)(u+v^2)\, +(5a-4) (1+u)\, v )}} 
\, = \,\,  idem(u',\, v')  \nonumber \\
&&  {{ du \cdot dv}
 \over {(v-1) \cdot ( (5a-4) (1+v)u+2(2a-1)(v+u^2) )}}
 \, =\,  \, idem(u', \, v')  \nonumber \\
&& {{ du \cdot dv} \over 
{ (v-u) \cdot ( 4(a-1) (1+uv)+(5a-2)(v+u) )}} \, = \, \, idem(u',\, v')
\end{eqnarray}
The second, and third, two-forms are obtained from the first one in (\ref{list})
by respectively
$(u,v)\rightarrow (v,u)$ for $b=2-2a$, and by
$(u,v) \rightarrow (u/v,1/v)$ with $a \rightarrow 1-a/2$, for $b=1-a/2$.
For the quadratic condition $C_0(a,b)=0$, the mapping preserves the following two-form,
{\em up to a minus sign}:
\begin{eqnarray}
&&   {{ du \cdot dv} \over { \rho(u, \, v) }}\,
= \, \, \, - \, {{ du' \cdot dv'} \over { \rho(u', \, v') }},
 \quad \quad \quad \quad \quad  \hbox{where :}\nonumber \\
\label{ellipt}
&& \rho(u, \, v)\, = \, \,\,\,
 (b-a)(a^2+b^2+3ab)\,  (1+u^2)\, v \nonumber \\
&&\quad \quad \quad -(2b+a)(a^2-b^2-ab)\, (1+v^2)\, u \label{list2} \\
&& \quad \, \,   \quad -(b+2a)(a^2-b^2+ab) \, (u^2+v^2)
+2(b-a)(2a+b)(a+2b) \, uv \nonumber
\end{eqnarray}
Note that $\,\rho(u, \, v)\, = \, 0$ is an elliptic curve.  
\vskip.2cm

Considering the 25 points $(a, \, b)$, listed in Appendix F,
for which the mapping is integrable, one can see that they all belong
to the codimension-one subvarieties
of the $(a, \, b)$ plane, where preserved meromorphic
two-forms are found, i.e. 
the curve $C_0(a,b)=0$ and/or the lines $b=a$, $b=2-2a$, $b=1-a/2$
(see Figure \ref{f:fig1}, lower left corner).

Furthermore, when these codimension-one subvarieties
intersect, the deduced $(a, \, b)$ points correspond to  integrability of
the mapping. 
The algebraic
invariants corresponding to these  integrability cases, can easily
be deduced from the fact that, at the intersection
of two curves among $\,C_0(a,b)=0$, and the lines $b=a$, $b=2-2a$, $b=1-a/2$,
{\em one necessarily has two simple two-forms preserved} (up to a sign).
Performing the ratio of two such two-forms one immediately
gets algebraic invariants of the integrable mapping.
See Appendix F for examples of algebraic invariants deduced, for
integrable points $(a, \, b)$, from 
ratio of two preserved two-forms.

\vskip .1cm
 {\bf Remark:} One may have the feeling that the exact results
on preserved meromorphic two-forms, or in the previous sections
on complexity reduction for (\ref{defKanti}), are consequences 
of the fact that we restricted ourselves to $\, \xi \, = \, a+b+c-1\, = \, 1$,
the condition for collineation $\, C$ to be involutive (yielding $\, K$ and $\, K^{-1}$ 
to be conjugate). This is not the case. We give in Appendix G
miscellaneous examples of exact results valid when this
involutive condition on $\, C$ is not verified ($\, \xi \, = \, a+b+c-1\, \ne \, 1$).

\vskip .1cm
 
It is tempting, after such an accumulation
of preserved two-forms, to see the previous results
(\ref{list}), (\ref{list2}) as a restriction to these codimension-one
subvarieties  in the ($a, \, b)$-plane,
of a general (conformally preserved) meromorphic 
two-form valid in the whole $(a, \, b)$-plane. In view of the
expressions of the two-forms for the three lines on one side, and the 
expression associated with the elliptic curve (\ref{ellipt})
on the other side, one could expect, at first sight, this
 meromorphic two-form to be quite involved.
 Using a "brute force" method we
 have tried to seek, systematically, for meromorphic two-forms
 $\, d\mu(u, \, v)=du \cdot dv \,/\rho(u, \, v)$, with
an algebraic (polynomial) covariant $\, \rho(u, \, v) $ in the form:
\begin{eqnarray}
\label{rho12}
\rho(u, \, v) \, =  \, \, \, \,
\sum_{i=0}^{n_1} \sum_{j=0}^{n_2} c_{ij}\, u^i \, v^j .
\end{eqnarray}
The existence of such a polynomial covariant curve is {\em ruled out}
up to $n_1=n_2=18$.
Formal calculations seem hopeless here, in particular if the final 
result is a non existence of such an algebraic covariant
 of (\ref{defKanti}) for generic $a, \, b$.
One needs to develop another approach that
 might be also valid to prove a ``no go''
result like the {\em non existence} of an {\em algebraic} covariant
$\, \rho(u, \,v) $, and beyond, the non existence of
a ``transcendental'' covariant
$\, \rho(u, \,v)$, corresponding to some  {\em analytic but not algebraic} 
curve\footnote[5]{Along this line, one should recall the occurrence of a {\em transcendental
invariant} for a birational mapping given by the ratio of
 products of simple Gamma functions, providing
an example of ``transcendental'' integrability (see equation (31), paragraph 7
of~\cite{Noether}, or 
equation (20), paragraph (8.3)
 in~\cite{classif}, or equation (3.3) in~\cite{SIDEV}).
}.

\section{Orbit of the critical set: algebraic curves versus chaotic sets}
\label{from}

When a preserved (resp. conformally preserved)  meromorphic 
two-form  $du.dv/\rho(u,\, v)$ exists,
one has the following fundamental relation (\ref{jac}) between the 
algebraic expression $\rho(u,\, v)$ and the Jacobian of transformation  $\,K $:
\begin{eqnarray}
\label{fundam}
\rho \left( K(u,\, v) \right) \,\, = \, \,\, \xi \cdot J(u,\, v) \cdot
\rho(u,\, v)
\end{eqnarray}
where $\, \xi $ is a constant. When $\, \xi \, = \, +1$ the two-form
 is preserved.
When there exists an integer $\, M$, such that $\, \xi^M \, = \, 1$,
the transformation $\, K^M$, instead of $\, K$,  preserves a   
two-form.  When $\, \xi \, \ne \, +1$
(for any $M$, $\, \xi^M \,\ne \, +1\,$),
it is just conformally preserved. 
Let us restrict the previous fundamental relation (\ref{fundam})
to a point $ (u,\, v)$ such that the Jacobian of transformation $\, K$
vanishes,  $\, J(u,\, v)\, = \, 0$. The fundamental relation 
(\ref{fundam}) necessarily yields for such a point:
\begin{eqnarray}
\label{fundam2}
  \rho \left( K(u,\, v) \right) \,\, = \, \,\, 0
\end{eqnarray}

For birational transformations, the images of the 
curves $\, J(u,\, v)\, = \, 0$
are  not  curves but {\em blow down into set of points}. For mapping 
(\ref{defKanti}), the vanishing condition 
$\, J(u,\, v)\, = \, 0$  splits into three curves 
$u=0$,$\, v=0$ and $u=-av/(v(a-1)+a)$. The image of these three curves
{\em blow down} into three points $(u^{(1)}, \, v^{(1)})$,
$(u^{(2)}, \, v^{(2)})$ and $(u^{(3)}, \, v^{(3)})$.
Being covariant, $ \rho(u,\, v)$ not only vanishes at these points
(i.e. $ \rho(u^{(i)},\, v^{(i)})=0$ for $i=1, \, 2, \, 3$),
but {\em also on their orbits}:
\begin{eqnarray}
\label{fundam3}
\rho \left( K^N(u^{(i)},\, v^{(i)}) \right) \,\, = \, \,\, 0, \qquad \quad \quad
 N \, = 1, \, 2, \, \cdots, \,i=1, \, 2, \, 3  
\end{eqnarray}
One can thus construct a (generically) {\em infinite set} of points on
$ \rho(u,\, v)=0$,
as orbits of such ``singled-out'' points
 $(u^{(i)},\, v^{(i)})$ and visualize them, whatever (the accumulation of) this set of 
points is (algebraic curves, transcendental analytical curves, chaotic 
set of points, ...). 

Before visualizing some orbits, let us underline that (\ref{fundam3})
means that the iterates of the critical set, also called {\em  post-critical set},
actually  cancel $\rho \left(u,\, v \right)$. These 
iterates are known in closed
forms for some subspaces. For instance, on the line $b=a$, the iterates
are given in Appendix E in terms of Chebyshev polynomials.
At these iterates $\left(u^{(i)}_N,\, v^{(i)}_N \right)$, with closed expressions,
one has $\rho \left(u^{(i)}_N,\, v^{(i)}_N \right)=0$.

The meromorphic two-forms found in Section (\ref{covar}) (see (\ref{list})),
actually correspond to situations such that the post-critical set (resp. the orbit of the
exceptional locus) has $\, \delta\, = \, 1$, closed expressions 
being available to describe all these points (Chebyshev polynomials, ...). 
The generic exponential growth (in the parameters) of the $\left(u^{(i)}_N,\, v^{(i)}_N \right)$ 
(namely $\, \delta \sim 2$), certainly excludes (even very involved)
algebraic
expressions (\ref{rho12}) for $\rho(u, \, v)$, but it may not exclude 
{\em transcendental analytical curves} (like the transcendental curves (31)
 in paragraph 7 of \cite{Noether}, or the transcendental curves (20)
 in \cite{classif},  which are orbits of a birational transformation exhibiting some
``transcendental'' integrability~\cite{Noether,SIDEV}.)

\subsection{Visualization of post-critical sets}
\label{visupost}
\begin{figure}
\psfig{file=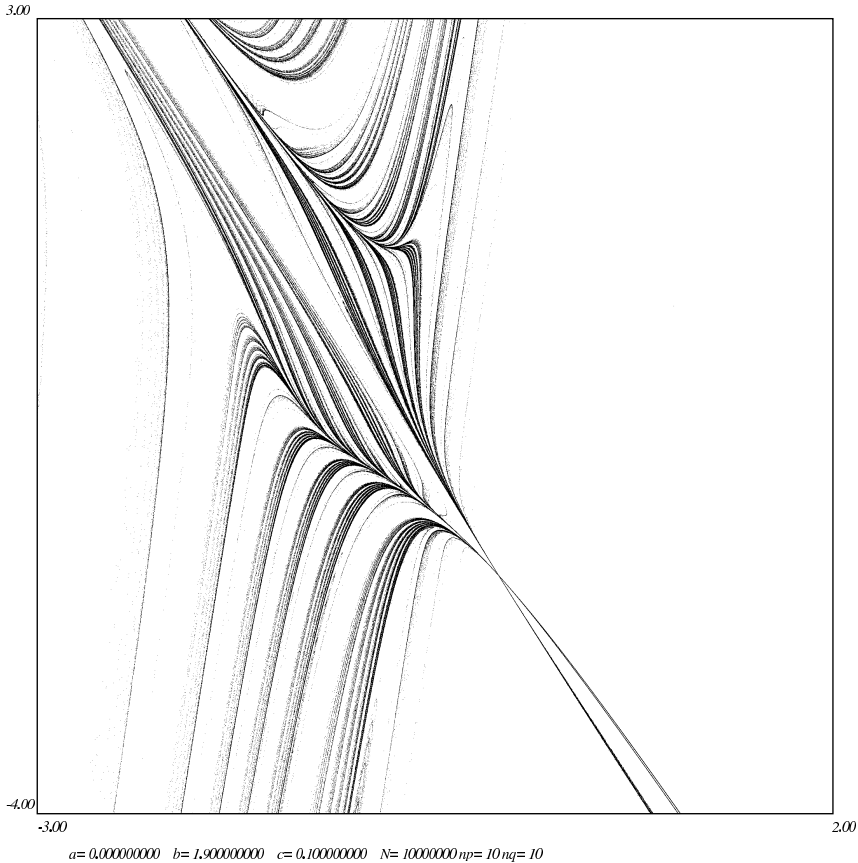,scale=0.6,bbllx=100bp,bblly=100bp,bburx=600bp,bbury=600bp}
\vskip .9 cm
\caption{Orbit of the critical set for $(a, \, b) \, = \, (0, \, 1.9)$}
\label{f:fig19}
\end{figure}

\begin{figure}
\psfig{file=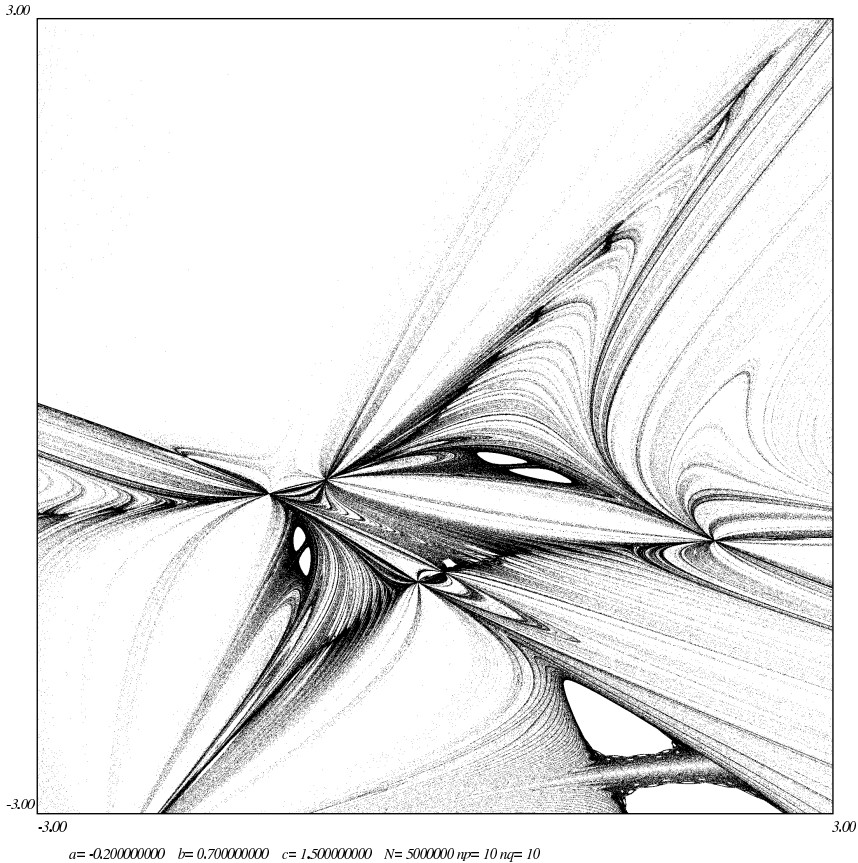,scale=0.6,bbllx=100bp,bblly=100bp,bburx=600bp,bbury=600bp}
\vskip .9 cm
\caption{Orbit of the critical set for $(a, \, b) \, = \, (-.2, \, .7)$}
\label{f:fig191}
\end{figure}

Let us visualize a few post-critical sets.
In the cases  where a meromorphic two-form is actually preserved (see (\ref{list})),
one easily  verifies that the orbit of $(u^{(1)}, \, v^{(1)})\, = \,(b/a, \, \, (1-a-b)/a)$,
actually yields  the (whole) covariant condition $\, \rho(u, \, v)\, = \, 0$ 
corresponding to the divisor of a meromorphic two-form 
when such a meromorphic two-form has been found. Of course if one 
performs iterations of other points (even very close) than
the singled out points as 
$\,(b/a, \, \, (1-a-b)/a)\,$, one will not get 
such algebraic covariant curve $\, \rho(u, \, v)\, = \, 0$, but 
more involved orbits. In contrast for parameters $(a, \, b)$ for which no
meromorphic two-form
was found, we see a drastically different situation, shown in
Figure \ref{f:fig19}, corresponding to the 
orbit of  $\, (b/a, \, \, (1-a-b)/a)$ (image by 
transformation $\, K$ of one of the vanishing conditions for the Jacobian)
for $(a, \, b) \, = \, (0, \, 1.9)$.

 This post-critical set (Figure \ref{f:fig19}) looks very much like a set of curves, 
a ``foliation'' of the $\, (u, \, v)$-plane.
Figure  \ref{f:fig191} shows the post-critical set corresponding to the
case
$(a, \, b) \, = \, (-.2, \, .7)$.
With these two orbits it is quite clear that this set of points
{\em cannot be} a
simple algebraic curve $\, \rho(u, \, v)\, = \, 0$. 

 At this step the ``true'' nature of this set of points
 is almost a ``metaphysical'' question: 
is it a transcendental analytical curve infinitely winding, is it a chaotic 
fractal-like set ... ?  In particular when one takes a larger frame for plotting the orbit, 
the set of points becomes more fuzzy, and it becomes more and more difficult,
to see if these points are organized in curves, like Figure \ref{f:fig19}
which
suggests an (infinite ...) accumulation of curves. 
We have encountered many times such a situation (see paragraph 5.1
and Figures 13 and 14 in~\cite{ab-an-bo-ma-2000}). A way to cope with the 
fuzzy appearance of the orbit when the points go to infinity, is to perform
 a change of variables (see paragraph 5 in~\cite{ab-an-bo-ma-2000}): 
$(u, \, v) $ $\, \rightarrow \, $$(1/u, \, 1/v) $. 
Again one has the impression to see some kind of ``foliation of curves''
for the previously fuzzy points, but the 
points that were seen in Figure \ref{f:fig19} as organized like
 a ``foliation'' of curves, have now 
(in some kind of ``push-pull game'') become fuzzy
 sets. One way to avoid this ``push-pull'' problem,
 and thus, ``see the global picture'' amounts to performing our plots 
in the variables $\, u_c  \,= \,  u/(1+u+u^2)$ and 
$\, v_c  \,= \,  v/(1+v+v^2)$. These variables are such
that any orbit of {\em real} points
 will be in the box $\,[ -1, \,1/3] \times [ -1, \,1/3]$.
This trick {\em ``compactifies''automatically our orbits}. 
\begin{figure}
\psfig{file=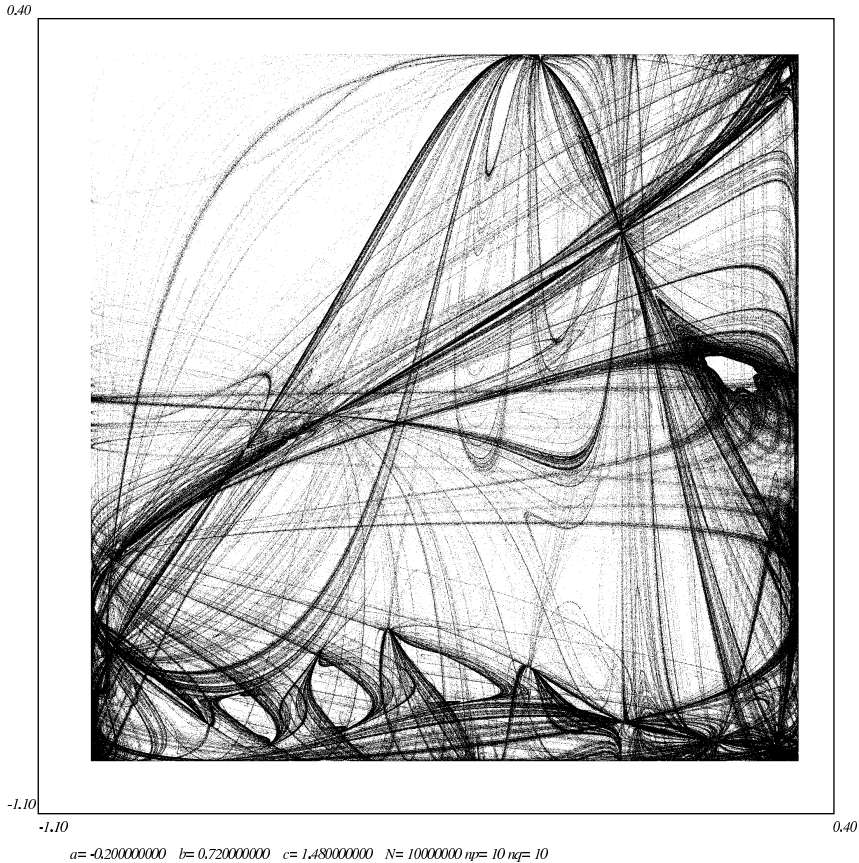,scale=0.6,bbllx=100bp,bblly=100bp,bburx=600bp,bbury=600bp}
\vskip .9 cm
\caption{The orbit of the critical set in the ``compact'' 
variables $\, u_c$ and  $\, v_c$, for $\, (a, \, b)  \, = \,( -.2, \,  .72)$}
\label{f:fig11}
\end{figure}
\begin{figure}
\psfig{file=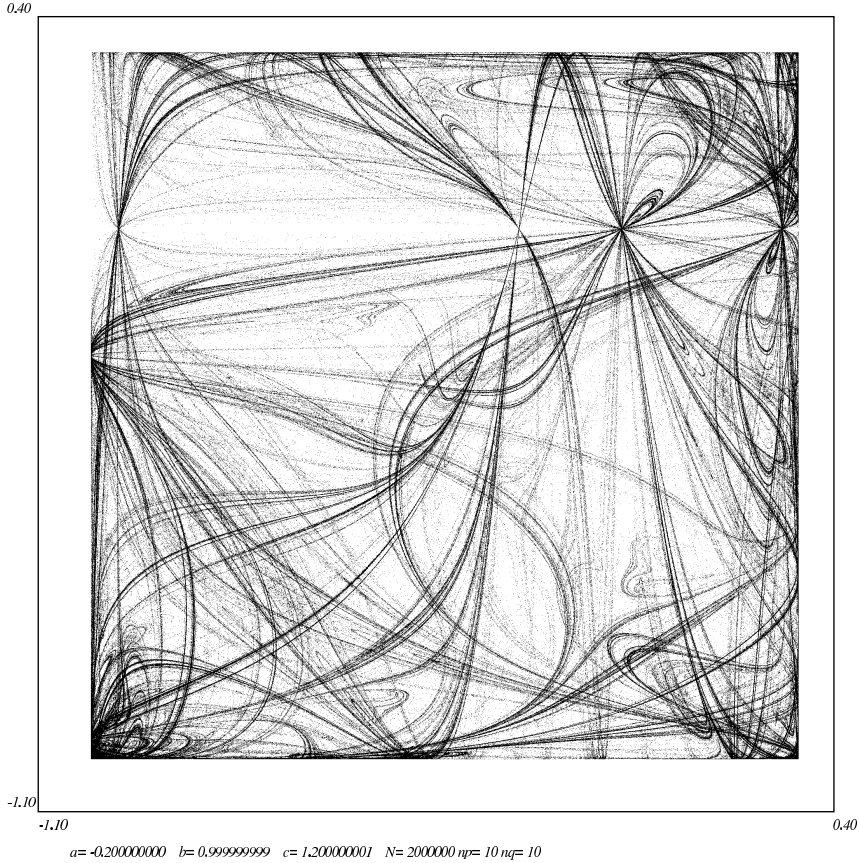,scale=0.6,bbllx=100bp,bblly=100bp,bburx=600bp,bbury=600bp}
\vskip .9 cm
\caption{The orbit of the critical set in the  ``compact'' 
variables $\, u_c$ and  $\, v_c$, for $\, (a, \, b)  \, = \,( -.2, \,  .999999999)$}
\label{f:fig7}
\end{figure}

Let us give two examples of such  ``compactified'' images of two 
orbits of $\, (b/a, \, \, (1-a-b)/a)$ (image by 
transformation $\, K$ of one of the vanishing conditions for the Jacobian).
For $\, (a, \, b)  \, = \,( -.2, \,  .72)$ one gets 
Figure \ref{f:fig11}, and for $\, (a, \, b)  \, = \,( -.2, \,  .999999999)$ one gets 
Figure \ref{f:fig7}.

In the situation where preserved meromorphic two-forms exist,
one sees that, even a very small deviation from the
 $\, (b/a, \, \, (1-a-b)/a)$ point 
(associated with the post-critical set), yields orbits that 
 look {\em quite different} from the algebraic covariant curve 
$\, \rho(u, \, v)\, = \, 0$. In contrast with this situation, we see, in the previous
cases where no preserved meromorphic two-forms exist, that a slight modification
of the  $\, (b/a, \, \, (1-a-b)/a)$ point (associated with the post-critical set)
yields  orbits which are {\em extremely similar} to the post-critical set 
of Figures (\ref{f:fig191})
or (\ref{f:fig11}), or (\ref{f:fig7}). These orbits are ``similar'',
 but {\em not converging towards 
this post critical set}. They are, roughly speaking, ``parallel''
to this post critical set. Therefore
 the orbit of the critical set
may be seen as a chaotic set, but it is a {\em  non attracting chaotic 
set} in contrast with the well-known strange attractors of H\'enon bi-polynomial
mappings~\cite{Bene,Vadim}.

\subsection{From preserved meromorphic two-forms and post-critical sets
back to fixed points}
\label{fromto}
Denoting $(u',\, v')$, $(u'',\, v'')$, ..., $\, (u^{(n)}, \, v^{(n)})$, 
the images of a point $ (u,\, v)$, by transformations
$\, K$,  $\, K^2$, ...,  $\, K^n$,
the preservation of a two-form yields
($ J[K^n](u, \, v)$ being the Jacobian of $\, K^n$):
\begin{eqnarray}
&& {{du \cdot dv } \over {\rho(u, \, v)}} \, \, = \, \, \,
  {{du' \cdot dv'} \over {\rho(u', \, v')}} \, \, = \, \,  \,
\cdots \, \, = \,  \,\,  {{du^{(n)} \cdot dv^{(n)} } \over {\rho(u^{(n)}, \, v^{(n)})}} \nonumber \\
&&J[K^n](u, \, v) \, \, = \, \, \,  {{ \rho(u^{(n)}, \, v^{(n)})} \over { \rho(u, \, v)}} 
\, \, = \,  \,\, {{ \rho(K^n(u, \, v))} \over { \rho(u, \, v)}} 
\end{eqnarray}
From the previous relation, it is tempting to deduce (a little bit too quickly ...) 
that the Jacobian of $\, K^n$ is equal to +1 when evaluated at the 
 fixed point $(u_f, v_f)$ of $\, K^n$: 
\begin{eqnarray}
\label{toquick}
J[K^n](u_f, \, v_f) \, = \, \, \, {{ \rho(K^n(u_f, \, v_f))} \over { \rho(u_f, \, v_f)}} 
 \, = \, \, +1
\end{eqnarray}
We actually found such strong results for (\ref{keps}), and for 
many other birational transformations
(when, for instance, we evaluated
precisely the number of $n$-cycles,
to get the dynamical zeta function~\cite{zeta}),
for
which a meromorphic two-form was actually preserved. In fact, even when a 
meromorphic two-form is preserved, relation (\ref{toquick}) (namely 
the Jacobian of $\, K^n$ evaluated at a fixed point of  $\, K^n$,
 is equal to +1), {\em may be ruled out} when the fixed 
points of $\, K^n$ {\em correspond to divisors of the two-form}.
If $\, \rho(u, \, v)$,
corresponding to a preserved meromorphic two-form,
is a rational expression $\, \rho(u, \, v)\, = \, P(u, \, v)/Q(u, \, v)$
($P(u, \, v)$ and $\, Q(u, \, v)$ are polynomials), the Jacobian of $\, K^n$,
evaluated at a fixed point
$\, (u_f, \, v_f)$ of  $\, K^n$, {\em can actually be different from} $\, +1$, if
$P(u_f, \, v_f)\, = \, \, 0$, or $\, Q(u_f, \, v_f)\, = \, \, 0$. 
Such ``non-standard'' fixed points of $\, K^n$ are such that $\, \rho(u_f, \, v_f) \, = \, \, 0$
(resp. $\, \rho(u_f, \, v_f) \, = \, \, \infty$), and of course, since $\rho(u, \, v)$
is typically a covariant of $\, K$ (see (\ref{fundam})), such that $\rho(u, \, v)$, evaluated at 
all their successive images by $\, K^N$ (for any $N$ integer), {\em vanishes}
(resp. is infinite):
\begin{eqnarray}
J[K^n](u_f, \, v_f)\, \ne \, +1, \quad 
\Rightarrow \quad \rho(K^N(u_f, \, v_f)) \, = \, \, 0 \,\quad  \hbox{(resp. $\,\infty$) } 
\end{eqnarray}
Performing orbits of such ``non-standard'' fixed points could thus be
seen as an {\em alternative way of visualization}
of  $\rho(u, \, v)$ (whatever its ``nature'' is: polynomial, rational expression, analytic 
expression, ... ), this alternative way being extremely similar to the 
one previously described, associated with the visualization of post-critical
sets.
Finding by formal calculations a very large 
accumulation of such  ``non-standard'' fixed points is not
sufficient to
prove the non-existence of meromorphic two-forms: one needs
to be sure that this accumulation of points cannot be localized on some
{\em unknown}
highly involved algebraic curve. It is well known that
proving  ``no-go'' theorems
is often much harder than proving theorems that simply require to exhibit
a structure.
However, as far as this difficulty to prove a
 non-existence is concerned, it can be seen as highly positive and effective,
 as far as simple
``down-to-earth'' visualization methods are concerned. In contrast with the
unique
post-critical set, we can consider orbits of a large (infinite) number of
such
``non-standard'' fixed  points of $\, K^n$.
The relation between the post-critical set
and such ``non-standard'' sets, is a very interesting one
that will be studied elsewhere. 

Let us just consider the birational transformation (\ref{defKanti})
for  $(a, \, b) =\, (-1/5, \, 1/2)$ (where no
meromorphic two-form has been found).
The primitive fixed points (cycles) and  the value
of the  Jacobian of $\, K^n$ at the corresponding fixed points,
 that we will denote $\, J$, are
 given~\footnote[5]{In these tables $\vert J \vert=1$ means that,
at the fixed point of $K^n$, the value of $J$  is complex and lying on the
 unit circle. Similarly, $J \ne 1$ means that $\, J$ is real, $\vert J \vert \ne 1$
that $\, J$ is not real.}
 in Table 1.
\vskip 5mm
\centerline{
\begin{tabular}{|l|l|l|l|l|l|l|l|l|}
\hline
$n$& 1& 2& 3& 4& 5& 6& 7& 8   \\ 
\hline
fix$(K^n)$& 4& 1& 2& 3& 6& 9& 18& 30   \\ 
$J=1$& 0& 1& 0& 1& 0& 3& 0& 6   \\ 
$\vert J \vert=1$& 2& 0& 2& 2& 4& 4& 6& 8   \\ 
$J \ne 1$& 2& 0& 0& 0& 2& 2& 4& 4   \\ 
$\vert J \vert \ne 1$& 0& 0& 0& 0& 0& 0& 8& 12   \\ 
\hline
\end{tabular}
}
\vskip 0.2cm
\textbf{Table 1:} Counting of primitive cycles
 for $a=-1/5,\,  b=1/2$. $\, J$ denotes $J[K^n](u_f, \, v_f)$.
\vskip 0.3cm
The number of cycles are in agreement with the Weil product expansion of 
the known (see (\ref{generic}))
exact expression of the dynamical zeta function:
\begin{eqnarray}
\label{weil}
&&\zeta(t) \, \, = \, \,     {{ 1} \over {\left( 1-2\,t \right)
 \left( 1-t \right) ^{2} }}\, \, = \, \,
{\frac {1}{ \left(1-t \right)^{4} \left(1-{t}^{2} \right)
 \left(1-{t}^{3} \right)^{2} \left(1-{t}^{4} \right)^{3}}} \, \times\nonumber \\
&&\quad \quad \quad  \quad \times \,
 {{1} \over {\left(1-{t}^{5} \right)^{6}
 \left(1-{t}^{6} \right)^{9} \left(1-{t}^{7}
 \right)^{18} \left( 1-{t}^{8} \right)^{30} \cdots }} 
\end{eqnarray}
To some extent, the situations where $\, J\, = \, -1$, or where $\, J$ 
is an $N$-th root of unity, can be ``recycled'' into a  $\, J \, = \, 1$ situation, 
replacing $\, K^n$ by $\, K^{2\, n}$ or $\, K^{N\, n}$.
However, we see on Table 1, the beginning
of a ``proliferation'' of  ``non-standard''points that cannot be reduced
to $\, J\, = \, -1$
or $\, J^N \, = 1$, strongly suggesting the non-existence of a 
meromorphic two-form. These enumerations have to be compared with the ones 
corresponding to   $(a, \, b) =\, (1/5, \, 1/5)$, for which a
meromorphic two-form is actually preserved. We still have the same sequence $\, 4,\, 1, \, 
 2, \, 3, \,  6, $$\,  9, \,  18,  \,  30 , \cdots$ of $\, n$-cycles, associated to the same
dynamical zeta function (\ref{weil}), however (except for the fixed points of order one), 
all the fixed points of order $\, n\, \ge 2$ are such
that $\, J \, = \, 1$~\footnote[1]{Recall that mappings (\ref{keps}) and (\ref{BiC2}) for
which a meromorphic two-form exists for generic values of the parameters,
are such that $\, J \, = \, 1$ for all the fixed points we have computed.}.

Along this line, let us consider mapping $K$ on 
curve $C_2^{22}(a,b)=0$, where, despite the complexity reduction,
no meromorphic two-form
has been  found. The calculations are performed
for the (generic) values $a=12/13,\,  b=-3/13$ (the number of
non generic $(a,b)$ on $C_2^{22}(a,b)$ is finite) and are given in Table 2.
\vskip 5mm
\centerline{
\begin{tabular}{|l|l|l|l|l|l|l|l|l|l|l|}
\hline
$n$&1&2&3&4&5&6&7&8&9&10   \\ 
\hline
fix$(K^n)$&4&1&2&2&4&5&10&15&26&42   \\ 
$J=1$&0&1&0&0&0&1&0&3&0&6   \\ 
$\vert J \vert=1$&2&0&0&0&2&2&0&2&4&2   \\ 
$J \ne 1$&2&0&2&2&2&2&6&6&10&14   \\ 
$\vert J \vert\ne1$&0&0&0&0&0&0&4&4&12&20   \\ 
\hline
\end{tabular}
}
\vskip 0.3cm
\textbf{Table 2:} Counting of primitive cycles
 for $(a, b)$ such that $C_2^{22}(a,b)=0$. $\, J$ denotes $J[K^n](u_f, \, v_f)$.

\vskip 5mm
The number of $ \, n$-cycles
are of course,
in agreement with the Weil product decomposition
of the exact dynamical zeta function (\ref{C2zeta}). 
We note the same proliferation of ``non-standard''
fixed points of order $\, n$, in agreement with the non-existence
of a preserved meromorphic two-form. This last result confirms 
what we saw several times, namely the {\em disconnection between 
the existence} (or non-existence) of a preserved meromorphic two-form
and (topological) {\em complexity reduction} for a mapping.

\subsection{Pull-back of the critical set : ``ante-critical sets'' versus post-critical sets}
\label{antecrit}

As far as visualization methods are concerned physicists, ``fortunately'', 
perform iterations without being conscious of the potential 
dangers: birational transformations have singularities and they 
may proliferate\footnote[2]{Are our numerical iterations
well-defined in some ``clean'' Zariski space, could ask mathematicians?} when performing iterations.
More precisely,
the critical set (vanishing conditions of the Jacobian) 
is a set of curves, whose images, by transformation $\, \, K$,
yield {\em points} and {\em not curves} (blow-down). 
 In contrast, the images  by transformation  $\, \, K^{-1}\, $ (resp.  $\, \, K^{-N}$) 
of these curves of the critical set (denoted in the following CS)
give {\em curves}: we do not have any blow-down with $\, \, K^{-1}$ 
(resp.  $\, \, K^{-N}$).
This infinite set of curves obtained by iterating
the critical set by $\, \, K^{-1}$, is such that, for some finite integer $\, N$,
the image of these curves by  $\, \, K^{N}$ {\em will blow down into points,
 after a finite number} $\, N$ of iterations. 
Let us call this set, for obvious reasons, ``ante-critical set''.
This ``ante-critical set'' is clearly
a ``dangerously singular'' set 
of points for the iteration of $\, K$. It is also a quite interesting
set from the singularity analysis viewpoint~\cite{bimero}. In particular, among these 
``dangerous points'' associated with the infinite set of curves $\, \Gamma_N \, = \, K^{-N}(CS)$,
some are singled-out (more ``singular'' ...): the points corresponding to 
intersections of {\em two} (or more) such curves 
$\, \Gamma_N$'s.  These singled-out points can, in fact, be obtained 
by some simple ``duality'' symmetries from the points of the post-critical set. 
Such ``ante-critical sets'', and their associated
 consequences on the birational transformations $\, K$,
 clearly require some further analysis that will be performed elsewhere.

\section{Lyapunov exponents 
and non-existence of meromorphic two-forms}
\label{detector}
\begin{figure}
\psfig{file=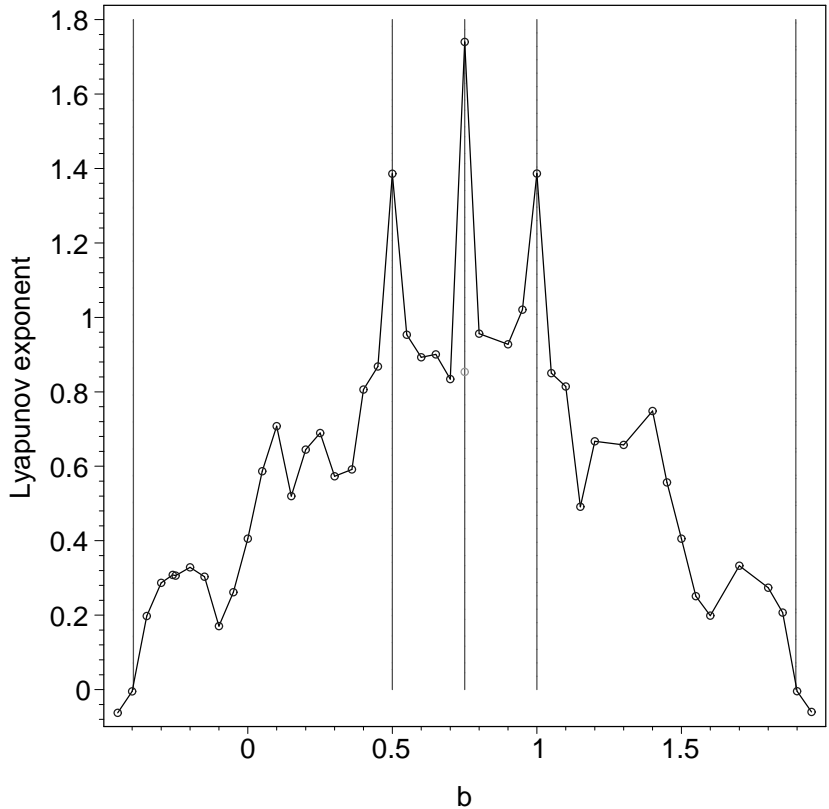}
\caption{Lyapunov exponents as a function of parameter $\, b$, for $\, a\, = \, 1/2$, the 
 initial point being $\, (u, \, v) \, = \, \, (2, \, 3)$.}
\label{f:fig4}
\end{figure}
\begin{figure}
\psfig{file=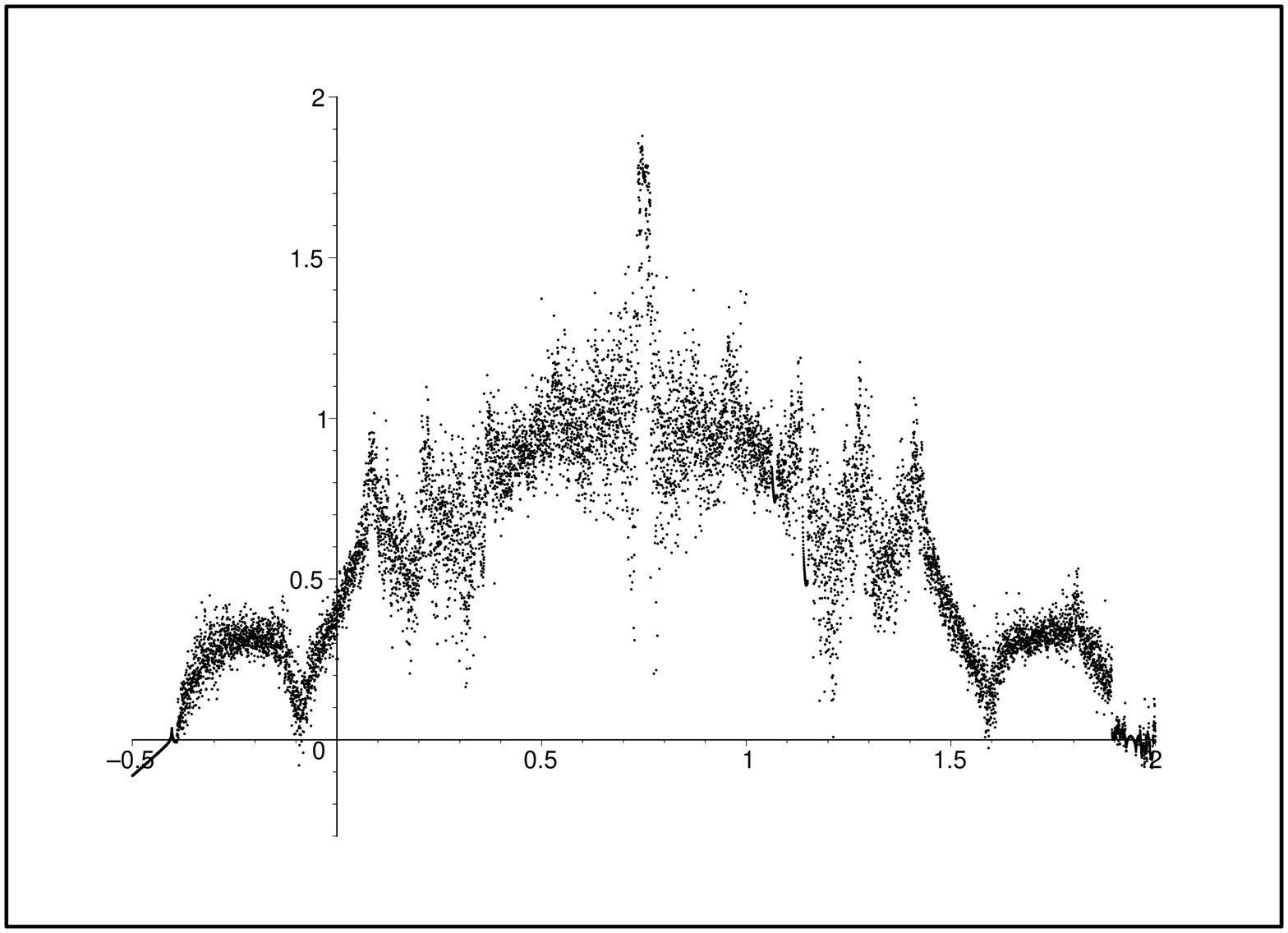,scale=0.65,angle=-90,bbllx=50bp,bblly=50bp,bburx=600bp,bbury=450bp}
\caption{Lyapunov exponents as a function of parameter $\, b$, for $\, a\, = \, 1/2$, the 
 initial point being the image of the critical set.}
\label{f:fig8}
\end{figure}

The previous simple visualization approach can be confirmed by some Lyapunov
exponents analysis.
Let us consider 
orbits of a given initial point 
(for instance $\, (u, \, v) \, = \, \, (2, \, 3)$)
under the iteration of birational transformation (\ref{defKanti})
for parameter $\, a$ fixed (for instance  $\, a\, = \, 1/2$), 
and for different values of the second parameter $\, b$, and let
us calculate the corresponding Lyapunov exponent. One thus gets 
the Lyapunov exponent (of what we can call a ``generic'' orbit) as a function 
of parameter $\, b$. This simple analysis is an 
easy down-to-earth way to detect
drastic complexity reductions, the complexity being not the 
topological complexity (like the topological entropy 
  or the growth rate complexity) but a less universal (more probabilistic)
complexity (like the metric entropy).

Figures (\ref{f:fig4}), (\ref{f:fig8}) show, quite clearly,  {\em non-zero and positive}
Lyapunov exponents, such results being apparently valid, not only for the
Lyapunov exponent
corresponding to our singled-out orbit, the post-critical set (see Figure (\ref{f:fig8})), but, also, for 
{\em every} orbit in the $(u, \, v)$-plane (see Figure (\ref{f:fig4})). With
 this scanning in the $\, b$ parameter
we encounter several times the singled-out cases where preserved meromorphic two-forms 
exist ($a\, = b$, $\, C_0(a, \, b) \, = \, 0$, ..., see (\ref{list})), and we see that these
specific points are singled-out on Figure (\ref{f:fig4}). 
If instead of performing the orbit of an arbitrary point ($(u, \, v) \, = \, \, (2, \, 3)$)
one calculates the Lyapunov exponent corresponding to the post-critical set
one finds similar results with a quite high volatility (a value of $\, b$
where the Lyapunov is a ``local'' maximum is quite close to a value where
 the Lyapunov is almost zero). 

In order to better understand this volatility, we have performed specific Lyapunov exponents calculations
restricted to the singled-out cases where preserved meromorphic two-forms 
exist ($a\, = b$, $\, C_0(a, \, b) \, = \, 0$, ..., see (\ref{list})).
In such cases we recover the situation we had~\cite{rearea}
 with birational mapping (\ref{keps}), namely 
the Lyapunov exponents are zero (or negative on the attractive fixed points)
for all the orbits we have calculated (the positive non-zero 
Lyapunov being possibly on some ``evanescent'' slim 
Cantor set~\cite{BedDill,DillBed}, see section (\ref{2-formversus}), 
that we have not been able to visualize numerically)
and the orbits always look like curves. It is clear that computer
 experiments like these, can hardly detect the slim
and subtle
Cantor sets corresponding to (wedge product) invariant measure described~\cite{BedDill,DillBed} by
Diller and Bedford
in such situations, associated with the narrow regions where non-zero positive Lyapunov 
could be found: within such (extensive) computer experiments  
we find, ``cum grano salis'',  that the  Lyapunov exponents are ``generically'' 
({\em as far as computer calculations are concerned ...}) zero. 

With this subtlety in mind, our computer experiments show clearly {\em non-zero positive} 
Lyapunov exponents when there is {\em no preserved} meromorphic two-form and
a total extinction of these Lyapunov exponents when such preserved meromorphic two-forms
take place.
 
The occurrence of non zero positive Lyapunov exponents for hyperbolic systems, or 
dynamical systems with strange attractors is well-known: this is not the 
situation we describe here.

\section{Conclusion}
\label{concl}

The birational transformations in $\, CP_2$,
introduced in section (\ref{antisto}),
which  generically do not preserve any meromorphic two-form,
are extremely similar to other birational transformations we previously studied~\cite{Noether},
which do preserve meromorphic two-forms. We note that these two sets of 
 birational transformations exhibit totally
 similar\footnote[3]{In fact identical
results: one gets the same family of polynomials controlling the
 complexity (see (\ref{shift}) or (\ref{familMNbis}) and
 compare with~\cite{Noether}).} results as far as 
 {\em topological complexity} is concerned (degree growth complexity, Arnold complexity
 and  topological entropy), but {\em drastically different numerical} results as far as a
 more ``probabilistic'' (ergodic) approach of
 dynamical systems is concerned (Lyapunov exponents).
With these examples we see that the existence, or non-existence,
of a preserved meromorphic two-form explains most of the (disturbing) 
apparent discrepancy, we saw, numerically,
between the topological and probabilistic approaches of such
dynamical systems.

The situation is as follows. When these birational mappings
preserve a mero-morphic two-form (conservative reversible case)
the (preliminary) results of  Diller and 
Bedford~\cite{BedDill,DillBed} on mapping (\ref{keps}) give a 
strong indication (at least in the region of the parameter
$\, \epsilon \, < \, 0$ )
that the regions where the chaos is concentrated, namely where 
the Lyapunov exponents are non-zero and positive, are quite evanescent,
corresponding to an extremely slim Cantor set
 associated with an invariant measure given by some wedge product.
This nice situation from a differential viewpoint (existence of a 
preserved two-form), is the unpleasant
one from the {\em computer experiments viewpoint}: it is extremely hard to see
the ``chaos'' (homoclinic tangles, Smale's horseshoe, ...)
from the analysis (visualization of the orbits, Lyapunov exponents
calculations, ...) of even very large sets of {\em real} orbits. 

On the contrary, when the birational mappings
do not preserve a meromorphic two-form, the regions 
where the Lyapunov exponents are non-zero, and {\em positive},  can, then,
clearly be seen on computer experiments. 

In conclusion, the existence, or non-existence, of preserved meromorphic
two-forms
has (curiously) no impact on the topological complexity of the 
mappings, but drastic consequences
on the {\em numerical appreciation} of the ``probabilistic'' (ergodic) complexity.

The introduction of the {\em post-critical set}, namely the orbit of
the points obtained by the blow-down of the curves corresponding to the
vanishing conditions of the Jacobian of the birational transformation,
thus emerges
as a fundamental concept, and tool (of topological and algebraic nature)
  to understand the 
probabilistic (and especially numerical) subtleties of the dynamics of such
 reversible~\cite{rob-qui-92,QuRo88} mappings.

\vskip .4cm 

\textbf{Acknowledgments}  We thank C. Favre 
for extremely useful 
comments on analytically stable birational transformations,
exceptional locus and indeterminacy locus, and its cohomology 
of curves approach of growth rate complexity.  We thank J-C. Angl\`es d'Auriac
and E. Bedford for many  discussions on birational transformations.
We also thank  J-P. Marco for interesting discussions on invariant 
measures.
(S. B) and (S. H) acknowledge partial support from PNR3.

\section{Appendix A: Algebraic geometry: singularities
 of curves as candidates for complexity reduction}
\label{appendC}
The conditions of reduced complexity give the points $(a,b)$ that
belong to the algebraic curves $C_N$. 
These algebraic curves are such that one has a reduced
 complexity for generic point $(a,b)$ on the curve.
However, singularities of these algebraic curves
(from a purely algebraic geometry viewpoint: local branches, ...)
can actually be seen to correspond to  points $(a, \, b)$ in the
parameter plane {\em yielding lower  complexities}
for the birational transformation $\, K$.

On each curve $C_N$, the spectrum of complexity at the singularities is
given by
\begin{eqnarray}
\label{formula}
1-2t+t^{p+2}\, =\,\, \, \,0, \qquad \quad p\, =\, 0,\, 1,\,  \cdots,\,  
N/2-2
\end{eqnarray}
For example, a generic point on the curve $C_8^{22}$, has the complexity growth
$\lambda=1.9980$. The singularities of this curve are non generic points
and have complexity growth $\lambda \simeq 1, 1.6180, 1.8392$ given by
(\ref{formula}) for $N=8$. The next curve $C_{10}^{22}$ with $\lambda \simeq 1.9995$,
will inherit the last three values and adds (since $p$ goes now to 3)
$\lambda \simeq 1.9275$. Note that for a given curve $C_N$, the largest value of
complexity growth reached by its singularities is given by $1-2t+t^{N/2}$.

Let us give the generating functions of the degrees $\, d_N$, and
genus $\, g_N$, of the successive  $C_N(a,\, b)=0$ algebraic curves.
Let us also introduce the generating function for $\, S_N$,
the number of singularities of the algebraic curves  $C_N$:
\begin{eqnarray}
d_C(t)  =  \sum_{n=1}^{\infty} d_{2\, n} \cdot t^{2\, n}, \quad 
g_C(t)  =  \sum_{n=1}^{\infty} g_{2\, n} \cdot t^{2\, n} , \quad 
s_C(t)  =  \sum_{n=1}^{\infty} S_{2\, n} \cdot t^{2\, n} \nonumber
\end{eqnarray}
They read respectively (for $C_N^{22}$ up to $N=12$):
\begin{eqnarray}
\label{dgCN}
&& d_C(t) \, = \,  \, \, \, \, \, 
2 \, t^2\, + \, 6 \, t^4 \, + \, 12 \, t^6 \, + \, 26 \, t^8 \, + \, 
48 \, t^{10} \, + \, \, 98 \, t^{12} \, + \, \cdots  \nonumber \\
&& g_C(t) \, = \, \, \, \, \, 
0 \, t^2\, + 5 \, t^4 \, + \, 20 \, t^6 \, + \, 73 \, t^8 \, + \, 
182 \, t^{10} \, + \, \, 491 \, t^{12} \, + \, \cdots  \nonumber\\
&& s_C(t) \,  = \, \, \, \, \, 
0 \, t^2\, + 5 \, t^4 \, + \, 15 \, t^6 \, + \, 31 \, t^8 \, + \, 
53 \, t^{10} \, + \, \, 113 \, t^{12} \, + \, \cdots  
\nonumber
\end{eqnarray}
The degrees  $\, d_N$, the genus $\, g_N$, and the  number of singularities $\, S_N$
clearly grow exponentially like
$\, \lambda^{2n}$ with  $\, \lambda \, < \,  2$.
We have no reason to believe that these three generating
 functions $\, d_C(t)$,  $\, g_C(t)$ and  $\, s_C(t)$,  
could be rational expressions. Similarly, their
 corresponding coefficients growth rates, $\, \lambda$,
have no reason, at first sight, to be  algebraic numbers. 

A singularity of an algebraic curve is characterized by  
the coordinates of the singularities in homogeneous 
variables, the multiplicity $m$, the delta invariant $\delta$ and
the number of local branches $r$.
In general $m \ge r$ and $\delta \ge m(m-1)/2$. The equality holds for 
all the singular points of $C_N$, however, as $N$
 increases, some points do not satisfy the equality.
These points are $(a=0, b=1)$, $(a=1, b=0)$, $(a=1, b=1)$ and $(a=0, b=0)$,
$(a=0,b=2)$, $(a=2, b=0)$.

\section{Appendix B: Computing  complexity growth
of points known in their floating forms }
\label{appendQ}

Let us show how to compute the complexity growth
of generic (algebraic) points on algebraic curves, and 
how to compute the complexity growth
of points known in their floating forms.

To compute the complexity growth for the parameters
$(a,b)$ belonging to a whole curve, e.g. $C(a,b)=0$, we fix $v$ (for
easy iteration), and we iterate up to order $N$.
We eliminate $b$ between the numerator of $u_N-X$ and the curve $C(a,b)=0$.
We can obtain factorizable polynomials $P_1 \cdot P_2 \cdots$
One counts the degree of $u$ in the polynomials depending on $X$, and  
discards the polynomials $P_i$ that contain only $u$.
Let us show how this works.
One considers the curve $C_2^{22}$ given in (\ref{C222}) and computes
the complexity for the parameters $a$ and $b$ such that $C_2^{22}(a,b)=0$. 
Let us fix $v$, and eliminate $b$ between 
$u_N-X$ and $C_2^{22}(a,b)$ ($u_N$ is the $N$-th iterated, one 
may take $v_N$ instead). One gets for the first four iterations
$P(X^2, u^2)$, $P(X^2, u^4)$, $P(X^2, u^8)$ and
$P(u^2) \cdot P(X^2, u^{14})$,
where $P(u^n)$, $P(X^n, u^p)$ denote polynomials in $X$ and $u$ with the 
shown degrees.
At step $4$, a polynomial in $u$ factorizes, which means that
the sequence of degrees in this case is $[1, 2, 4, 8, 14, \cdots]$
instead of the generic $[1, 2, 4, 8, 16, \cdots \, ]$. 

The degrees of the curves grow as the iteration proceeds, we may need, 
then, to compute the growth complexity for points in the $(a,b)$-plane
{\em only known in their floating form}. We introduce a float numerical method
that deals with these points obtained as roots of polynomials of degree
greater than five.
The method
starts with the parameters in their floating forms. The iteration
proceeds to order $N$, where one solves the numerator, and the denominator,
of the variable (say) $u_N$. We take away the common roots and so on.
The computation is controlled by the number of digits used. The
computation with the float numeric method is carried out on the
homogeneous variables. Let us show how the method works.
The parameters $a$ and $b$ are fixed, and known, as floating numbers (with the desired 
number of digits). The iteration proceeds as (in the homogeneous variables 
$(x,y,t)$, where we may fix the starting values of $y$ and $t$):
\begin{eqnarray}
x & \rightarrow &\,  x_1=P_1^x(x)\, \rightarrow\,
 x_2=P_2^x(x)\, \rightarrow \,\cdots \nonumber \\
y & \rightarrow & \,y_1=P_1^y(x)\, \rightarrow \,
y_2=P_2^y(x)\, \rightarrow \,\cdots \nonumber \\
t & \rightarrow &\, t_1=P_1^t(x) \,\rightarrow \,
t_2=P_2^t(x)\, \rightarrow\, \cdots \nonumber 
\end{eqnarray}
At each step, solving in  float each expression, amounts to writing:
\begin{eqnarray}
P_i^x(x)= \Pi_{j=1}^{n_1} (x-\tilde{x}_j), \, \,
P_i^y(x)= \Pi_{j=1}^{n_2} (x-\tilde{x}_j), \,  \, 
P_i^t(x)= \Pi_{j=1}^{n_3} (x-\tilde{x}_j), \nonumber 
\end{eqnarray}
The common (up to the fixed accuracy) terms $(x-\tilde{x}_j)$ between
$P_i^x(x)$, $P_i^y(x)$ and $P_i^t(x)$ are taken away and
the degree of, e.g., $P_i^x(x)$  {\em is counted according to this 
reduction}.

\section{Appendix C: Degree growth complexity and the ``arrow of time''}
\label{appendZ}
Let us consider (after V. Guedj and N. Sibony~\cite{Sibony,Sibony2}) 
the following bi-polynomial transformation:
\begin{eqnarray}
K(x, \, y, \, z)\,\, = \, \, \, \, \Bigl(z
, \, \, y  -z^{d}, \, \,x\, + \, y^2 -2 \, y \, z^{d} \Bigr)\nonumber
\end{eqnarray}
Its inverse reads: 
\begin{eqnarray}
K^{-1}(x,y,z)\, =\, \, \Bigl(z-y^2+x^{2d}, \, y+x^d, \, x  \Bigr)\nonumber
\end{eqnarray}
Written in the homogeneous variables $u,v,w,t$,  transformation $\, K$, and its inverse,
become:
\begin{eqnarray}
&&K(u,v,w,t)\, =\,\, \Bigl(wt^d, \, vt^d-t w^d, \, ut^d+v^2t^{d-1}-2vw^d, \, t^{d+1}  \Bigr)
\nonumber \\
&&K^{-1}(u,v,w,t)\,\, = \,\,\nonumber \\
&& \qquad \Bigl(wt^{2d-1}-v^2t^{2d-2}+u^{2d}, \,t^d \, ( vt^{d-1}+u^d),
\, ut^{2d-1}-2vw^d, \, t^{2d}  \Bigr) \nonumber
\end{eqnarray}

Fixing $d=1$, for heuristic reason, 
the successive degrees of $K^n(u,v,w,t)$ read
\begin{eqnarray}
deg_u\,=\,\,deg_v\,=\,\,deg_w\,=\,\,deg_t\,=\,\,
[2,3,5,8,13,21,34,55,\,  \cdots\, ]\nonumber
\end{eqnarray}
giving the degree generating function
\begin{eqnarray}
G(K)(t)\,= \,\,{\frac{t\cdot (t+2)}{1-t-t^2}}\nonumber
\end{eqnarray}
while the  successive degrees of $(K^{-1})^n(u,v,w,t)$ read
\begin{eqnarray}
deg_u\,=\,\,deg_v\,=\,\,deg_w\,=\,\,deg_t\,=\,\,
[2,4,8,16,32,64,\,  \cdots\, ]\nonumber
\end{eqnarray}
and give the degree generating function:
\begin{eqnarray}
\label{GKmoinsun}
G(K^{-1})(t)\,=\,\, {\frac{2t}{1-2t}}\nonumber
\end{eqnarray}
Transformation $\, K$ has clearly a golden number complexity 
different, and smaller, than the complexity $\, \lambda \, = \, 2$ of its inverse.

\section{Appendix D: A transcendental zeta function ?}
\label{appendF}

In this appendix, we consider the dynamical zeta function for the parameters ($a,b$) on
$\, C_0(a, \, b)\, = \, 0$. This is a bit subtle since the number
of fixed points for $\, K^2$ (and thus $\, K^{2N}$) is infinite (a 
whole curve (\ref{list2}) is a curve of fixed points of
 order two). Apparently one does not seem to have
even primitive cycles (except the infinite number of two-cycles). 
Introducing the zeta functions as usual by 
 the infinite Weil product~\cite{zeta}
on the cycles, avoiding the two-cycles and taking 
into account just the
odd primitive cycles one could write:
\begin{eqnarray}
\label{what}
 1/\zeta(t) \,\, = \, \,\,
 ( 1-t )^{4} ( 1-{t}^{3})^{2} 
( 1-{t}^{5})^{6}
 (1-{t}^{7})^{18} (1-{t}^{9})^{56}
 (1-{t}^{11})^{186} \cdots \nonumber
\end{eqnarray}

Recalling the ``generic'' expression (\ref{generic}), 
this expression $\zeta(t)$ is such that
$\zeta(t)\, \zeta_g(-t)=\zeta(-t)\, \zeta_g(t)$,
and verifies the following functional relation 
\begin{eqnarray}
 \zeta(t) \, = \, \, \,\, {{1+t} \over {1-t} } \cdot
 \Bigl({{1+2t} \over {1-2t} } \Bigr)^{1/2} \cdot  \zeta(t^2)^{1/2}\nonumber
\end{eqnarray}
yielding an infinite product expression for $\,  \zeta(t)$:
\begin{eqnarray}
\label{infprod}
\zeta(t) \, = \, \, {{1+t} \over {1-t} } \cdot
\prod_{i=0}^{\infty} 
\Bigl({{(1+2\, t^{2^i}) (1+t^{2^{i+1}})} 
\over {(1-2\, t^{2^i})(1-t^{2^{i+1}})} } \Bigr)^{1/2^{i+1}} \nonumber
\end{eqnarray}

For $n$ as upper limit of the above infinite product, the expansion
is valid up to $t^{2^{n+1}-1}$.
The ratio of the coefficients of (for example)
$t^{1023}$ with $t^{1022}$ gives
$\, \lambda \, \simeq \,1.9989099 $, 
in agreement with a complexity  $\, \lambda \, = \, 2$, but 
{\em with a dynamical zeta function that {\em is not} a rational expression, 
but some ``transcendental'' expression}.

Of course one can always imagine that the ``true'' dynamical zeta function
requires the calculation of all the ``multiplicities'' of Fulton's 
intersection theory~\cite{Fulton}, and that this very zeta function
is actually rational ...

\section{Appendix E: The mapping on the lines $b= \pm a$ and $C_0(a,b)=0$}
\label{appendB}

Along  the line $b=a$ (and similarly on its equivalents obtained 
by the actions of $P$ and $T$),
the growth of the degrees of the parameter $a$ in the iterates
of the vanishing conditions of the Jacobian is polynomial ($\delta=1$). 
One, then, expects the iterates to be
given in closed forms. This is indeed the case as can be seen below.
The iterates $K^n(V_1)$  are given by
\begin{eqnarray}
&&K^n(V_1) = \Bigl(u_n, \,\, v_n \Bigr)\qquad \hbox { with: } \qquad 
\sigma_1 = {\frac{3a^2-4a+2}{2(2a-1)}} \nonumber \\
&&u_n  = \, \,  {\frac{2(2a-1)\, T_{n}(\sigma_1)+a(5a-4)\,U_{n-1}(\sigma_1)-2(2a-1)}
{2(2a-1)\,T_{n}(\sigma_1)+a(5a-4)\,U_{n-1}(\sigma_1)+2(2a-1) }}     
\nonumber \\
&&v_{2n}  = \, \,  
{\frac{-2(2a-1)(5a-4)\,T_{n}(\sigma_1)-3a(3a-2)(a-2)\,U_{n-1}(\sigma_1)}
{4(2a-1)^2\,T_{n}(\sigma_1)}}     \nonumber \\
&&v_{2n-1}  =  \, \, 
{\frac{2(2a-1)(a^2+2a-2)\,T_{n}(\sigma_1)-a(3a-2)(a-2)^2\,U_{n-1}(\sigma_1)}
{-2a\,(2a-1)^2\,T_{n}(\sigma_1)+a(a-2)(3a-2)(2a-1)\,U_{n-1}(\sigma_1)}}     
\nonumber
\end{eqnarray}
where $T_n, U_n$ are Chebyshev polynomials of order $n$ of, 
respectively, first and second kind. 

We have very similar results for  the iterates $K^n(V_3)$. The iterates 
$K^n(V_2)$  are quite simple and read:
\begin{eqnarray}
K^n(V_2)\, = \, \, \Bigl(1, \,\, {\frac{2(2a-1)U_{n-1}(\sigma_2)}
{2T_n(\sigma_2)-(5a-4)U_{n-1}(\sigma_2) }} \Bigr), \quad \, 
\sigma_2  =  (3a-4)/2 \nonumber
\end{eqnarray}

For $(a, \, b)$ parameters such that $C_0(a,b)=0$, the iterates of the vanishing
conditions of the Jacobian are also given in closed forms and the growth
of the degrees of the parameters is {\em polynomial} ($\delta=1$).

Note that one finds similar results along the line $b=-a$
(and similarly on its equivalents obtained 
by the actions of $P$ and $T$) the growth of the degrees of the parameter $a$ in the iterates
of the vanishing conditions  of the Jacobian is also polynomial ($\delta=1$)
for $K^n(V_2)$. However, it is non-polynomial for $K^n(V_1)$ and $K^n(V_3)$
($1 < \delta \le 2$). The iterates $K^n(V_1)$ 
and $K^n(V_3)$ are not given as closed expressions.
Those $K^n(V_2)$  are given by:
\begin{eqnarray}
K^n(V_2) =\,   \Bigl(-1, \,\, {\frac{2U_{n-1}(\sigma)}
{2T_n(\sigma)+aU_{n-1}(\sigma) }}  \Bigr), \quad \quad 
\quad \sigma =  a/2 \nonumber 
\end{eqnarray}
$K^n(V_2) \in I_2$ gives the points 
where the curves $C_N^{ij}$ are tangent to the line $b=-a$.

\section{Appendix F: Cases of integrability}

The points $(a,b)$ for which the mapping $K$ defined in (\ref{defKanti})
is integrable are shown in
Figure \ref{f:fig1} (lower left corner). These points are lying on the lines (solid lines) $b=a$,
$b=2-2a$ and $b=1-a/2$, and on the curve $C_0(a,b)=0$ (ellipse). The dashed 
lines in Figure \ref{f:fig1} (lower left corner) are $b=-a$, $b=2$ and $a=2$.

On the lines $b=a$ and $b=2-2a$, the integrable cases are:
\begin{eqnarray}
\label{intaeqb}
 a\,=\,\,0, \,\,\,\,\, {{1}\over{3}}, \,\,\,\,\,
 1-{{1}\over{\sqrt{3}}},  \,\,\,\,\,
{{2}\over{3}}, \,\,\, \,\,1, \,\,\,\,\, {{4}\over{3}}, 
 \,\,\,\,\, 1+{{1}\over{\sqrt{3}}}
\end{eqnarray}
On line $b=1-a/2$, the integrable cases, obtained by applying $T \cdot P$,
are given by $(2-2a, a)$ from (\ref{intaeqb}).
The point $(a=2/3, b=2/3)$ is common to three lines and
corresponds to a matrix of the stochastic form (\ref{C2})
and the ``antistochastic'' form (transpose)  in the same time. 

From these 19 points $(a,b)$, the following six are also on the curve $C_0(a,b)=0$
\begin{eqnarray}
\label{C0intab}
 (-2/3, 4/3), \quad (4/3, -2/3), \quad (4/3, 4/3), \quad (0,0), \quad 
(0,2), \quad (2,0)  \nonumber 
\end{eqnarray}
The curve $C_0(a,b)=0$ has six other integrable cases:
\begin{eqnarray}
\label{C0intC0}
\Bigl({{1 \, \pm \sqrt{5}}\over{2}}, 1 \Bigr), \,\, \quad 
\Bigl({{1\, \pm \sqrt{5}}\over{2}}, {{1 \, \mp \sqrt{5}}\over{2}} 
\Bigr), \,\,\quad 
\Bigl(1, {{1\, \pm \sqrt{5}}\over{2}} \Bigr), \,\,
\end{eqnarray}
One has a total of 25 values of $(a,b)$ for which the mapping $\, K$ is 
integrable.

The integrable points common to $C_0(a,b)=0$ and the lines $b=a$,
$b=2-2a$, and $b=1-a/2$,
can be understood from the
existence of the two preserved two-forms.
Let us consider, for instance, the point
 $\, (a, \, b) \, = \, (0, \, 2)$ intersection of $\, C_0(a, b) \, = \, 0$
 and $\, b \, = 2-2\, a$.
Transformation $\, K$ for $\, (a, \, b) \, = \, (0, \, 2)$ 
preserves
 two two-forms respectively associated with $\, b\, = \, 2 -\, 2\, a$ in (\ref{list}),
and $\, C_0(a, b) \, = \, 0$ (see  (\ref{list2})), namely:
\begin{eqnarray}
\label{bothtwoform}
{{ du \cdot dv} \over { (1-v) \cdot ((v+{u}^{2})\, +2\,u (1+v)) }}, \quad \quad 
{{ du \cdot dv} \over { (1+v) \cdot ((v+{u}^{2})\, +2\,u (1+v)) }} \nonumber
\end{eqnarray}
corresponding to the fact that  $\, K$ has (up to a sign)
 $\, Inv \, = \, \, (1+v)/(1-v)$,  as an invariant. This is indeed the case since:
\begin{eqnarray}
\label{K2homo}
K^2(u, \, v) \, = \, \, \, \Bigl( -{{ (4 \, +7\,v\, +4\,{v}^{2}) \cdot u\,
 +2\,v \left( 1+v \right) }
 \over { 2\, (1+v) \cdot u\, + \, v }}
, \, \,v \Bigr)
\end{eqnarray}
We have similar results for the two other integrable points
 $\, (a, \, b) \, = \, (0, \, 0)$ and  $\, (a, \, b) \, = \, (2, \, 0)$. They 
also correspond to $\, K^2$ being a homographic
 transformation ($(a, \, b) \, = \, (0, \, 0)$
preserves the $\, u$ coordinate, and $\, (a, \, b) \, = \, (2, \, 0)$
 preserves the ratio $\, u/v$).
 Note that for the point $\, (a, \, b) \, = \, \, (1, \, 1)$, as well
 as $\, (1, \, 0)$ and $(0, \, 1)$,
the mapping  $\, K$ is of order six,
$\, K^6 \, = \, identity$.

The mapping $\, K$, 
for the integrable point  $\, (a, \, b) \, = \, \, (4/3, \, -2/3)$
  preserves two  two-forms:
\begin{eqnarray}
{{du \cdot dv} \over 
{\left( v-1 \right)  \left( 4\,u(1+v)+5\,(v+u^2) \right) }}, 
\qquad  \qquad 
  {{du \cdot dv} \over {
\left( v-1 \right)  \left( v-  {u}^{2} \right) }} \nonumber
\end{eqnarray}
their ratio giving the algebraic $\, K$-invariant (up to sign):
\begin{eqnarray}
Inv \, = \, \, 
{\frac {  v-{u}^{2} }
{ 4\,u(1+v)+5\,(v+u^2) }}
\end{eqnarray}

\section{Appendix G: miscellaneous exact results for $\, \xi \, = \, a+b+c-1\, \ne \, 1$}
\label{appendX}

Let us provide here a set of exact results, structures (existence of meromorphic two-forms ...) 
 valid in the more general framework where 
 $ c \, \ne \, 2-a-b$ ($\, K^N$ and  $\, K^{-N}$
are no longer conjugate).

When  $ c \, \ne  \, 2-a-b$,
the resultant in $\, u$ of the two conditions of
 order two of birational transformation (\ref{defKanti}),
namely $\, K^2(u, \, v) \, = \, \, (u, \, v)$,
yields the following condition (reducing to condition
 $\, C_0(a, \, b)\,=\,0$ previously 
written, when  $ c \, =  \, 2-a-b$):
\begin{eqnarray}
\label{newcond}
a\,b\,+b\,c\,+c\,a  \, =  \,\, \, \,0
\end{eqnarray}
associated with the (quite symmetric) homogeneous $\, K$-covariant ($K^2$-invariant) 
in the $(x, \, y, \, t)$ homogeneous variables:
\begin{eqnarray}
\label{covorder2}
&&cov(x, \, y, \, t) \, = \, \, 
\,bc \cdot t \left( {y}^{2}-{x}^{2} \right) +\,ac \cdot x \left( {t}^{2}-{y}^{2}
 \right) +\,a b\cdot y \left( {x}^{2}-{t}^{2} \right)
\nonumber \\
&&\quad +\, \left( yt+xt+xy \right)  \left(  \left( c-b \right) bc\cdot t\, + \left( b-
a \right) a b \cdot y \, + \, ac \cdot\left( a-c \right)\cdot x \right)\nonumber 
\end{eqnarray}

One easily finds that, restricted to (\ref{newcond}), the following
meromorphic two-form is preserved up to a minus sign:
\begin{eqnarray}
 {{ dx' \cdot dy' } \over { cov(x', \, y', \,1)}} \,\, \,  =  \,\,\, \,  (-1) \times  
{{ dx \cdot dy } \over { cov(x, \, y, _,1)}}
\end{eqnarray}

\subsection{For $\, b \, = \, c$, 
when $ c \, \ne  \, 2-a-b$: more two-forms. }

Keeping in mind the simple results (\ref{list})
for meromorphic two-forms (\ref{rho12}), let us restrict to the case
where the $\, K$-covariant $\, \rho(u, \, v)$
in a meromorphic two-form like (\ref{rho12}), is a {\em polynomial}, instead 
of a rational (algebraic, ...)
expression. 
Let us remark that when $\, c\, = \, b$ 
but $ c \, \ne  \, 2-a-b$, $\, u-v$ is a covariant of transformation $\, K$
with cofactor $\, 1/((a-1) \, u\, v\,  +a (u+v))$. Recalling expression (\ref{jact}) 
of the Jacobian of (\ref{defKanti}), it becomes quite natural, when $\, b=c$, 
 to make an ``ansatz'' 
seeking for covariant polynomials $\, \rho(u, \, v)$ of the form $\,\rho(u, \, v)\, 
= \, (u-v) \cdot Q(u, \, v)$, where  $\, Q(u, \, v)$ 
will be a  $\, K$-covariant quadratic polynomial
with cofactor  $\, \xi \cdot u\, v/((a-1) \, u\, v\,  +a (u+v))^2$. After
 some calculations, one finds
that the quadratic polynomial $Q(u, \, v)$ must be of the form:
\begin{eqnarray}
Q(u, \, v) \, = \,\,\,\,
A \, {a}^{2}\,uv\,\,  + \,B \,{a}^{2}\cdot  (u+v)
 -a \, (2\,b-1)\, B\, -b
\, (b-1)\,  A \nonumber
\end{eqnarray}
the $(a, \, b)$ parameters being necessarily such that :
\begin{eqnarray}
&&\left( b-a \right)  \left( a+b+c-2 \right) 
 \left( ab+bc+ac \right)
\, \, = \, \, 0 \qquad \hbox{and :} \nonumber \\
&&{a} \left( b+c-1 \right)  \left( b+a \right)  
\left( {b}^{2}+ab+{a}^{2}-a-b-c+1 \right) \nonumber 
\, \, = \, \, 0 
\end{eqnarray}
$\, \bullet \, $ Conditions $\, b \, = \, c \, = \, -a\, $
 yields $A \, = \, B$, and the conformally preserved two-form
reads ($\xi=a+b+c-1$):
\begin{eqnarray}
&&{{ du' \cdot dv'} \over { (u'-v') \cdot (u'+1)  \cdot (v'+1) }}
\,   = \, \,\,\, \, \xi \cdot 
{{ du \cdot dv} \over { (u-v) \cdot (u+1)  \cdot (v+1) }}
 \nonumber 
\end{eqnarray}
$\, \bullet \, $ Conditions $\, b \, = \, c \, = \, a\, $
 yields the conformally preserved two-form:
\begin{eqnarray}
\label{aegbegc}
{{ du' \cdot dv'} \over {(u'-v') \cdot (u'-1) \cdot (v'-1)}}\,\, = \, \, \,\, \,
\xi \cdot 
{{ du \cdot dv} \over {(u-v) \cdot (u-1) \cdot (v-1)}} \nonumber
\end{eqnarray}

\subsection{For  $ c \, \ne  \, 2-a-b$: more complexity reductions }
\label{miscell}
Condition $K^2(V_2) \in \, I_2$ amounts to writing
\begin{eqnarray}
  K^2\Bigl(  u, \, \,0 \Bigr)  \,\,
 = \,\,K \Bigl( {{b} \over {a}} , \, \,{{ c-1} \over {a}} \Bigr) 
 \, = \, \, \Bigl( {{a} \over {b}} , \, \,{{a} \over {1-a-b}} \Bigr)
 \nonumber
\end{eqnarray}
which yields several algebraic curves,  in particular the rational curve 
 $\, (c, \, b) \, =$ 
$\, \, (1+1/2\,{a}^{2}, \,-a)$, for which one can verify 
that a reduction of the degree
growth rate complexity $\, \lambda \, \simeq \,1.839 $ takes place. The 
degree generating function reads:
\begin{eqnarray}
G_{b=-a,c=1+1/2\,{a}^{2}} \,\,  = \, \, \,
 {{1} \over { 1-t-t^2-t^3}} \,\,  = \,\,  \,
{{ 1-t} \over { 1-2\, t+t^4}}
\nonumber 
\end{eqnarray}
Similarly   $K^4(V_2) \in \, I_2$ yields several algebraic curves, in particular the rational curve 
 $\, (c, \, b) \, =$ $\, \, (1+a^2/3, \,-a)$, for which one can verify 
 a reduction of the degree
growth rate complexity $\, \lambda \, \simeq \,1.965 $, the degree generating function reading:
\begin{eqnarray}
&&G_{a=-b,c= 1+a^2/3} \, = \, \, \, {{1-t} \over {1\, -2\,t\, +t^6}}  \, = \, \, \, 
1+t+2\,{t}^{2}+4\,{t}^{3} \nonumber \\
&& \quad  \quad \quad \quad +8\,{t}^{4}+16\,{t}^{5}+31\,{t}^{6}+61\,{t}^{
7}+120\,{t}^{8}+236\,{t}^{9} \, + \cdots \, \, \,  \nonumber
\end{eqnarray}

\vskip .2cm 

This is just a set of results for $\, \xi\, \ne 1$, among many 
others that can be easily established.

\end{document}